\documentclass[reqno,a4paper,12pt]{amsart}
\usepackage[utf8]{inputenc}
\usepackage[normalem]{ulem}
\usepackage{amssymb,amsmath,epsfig,graphics,psfrag,graphicx,color,cite,subfigure,arydshln,comment}

\usepackage[colorlinks=true,breaklinks=true,linkcolor=lightgreen,citecolor=lightblue]{hyperref}

\definecolor{lightblue}{rgb}{0.22,0.45,0.70}
\definecolor{cgray}{rgb}{0.7,0.7,0.7}

\definecolor{lightgreen}{rgb}{0.22,0.55,0.20}

\numberwithin{figure}{section}
\numberwithin{table}{section}
\numberwithin{equation}{section}

\DeclareMathAlphabet\mathbit
    \encodingdefault\rmdefault\bfdefault\itdefault
\DeclareOldFontCommand{\bi}{\normalfont\bfseries\itshape}{\mathbit}

\newcommand\bD{\boldsymbol{D}}
\newcommand\bU{\boldsymbol{U}}
\newcommand\bd{\boldsymbol{d}}
\newcommand\bu{\boldsymbol{u}}
\newcommand\bv{\boldsymbol{v}}

\newcommand\bx{\boldsymbol{x}}
\newcommand\bX{\boldsymbol{X}}

\newcommand\nn{\boldsymbol{n}}

\newcommand\bt{\boldsymbol{t}}
\newcommand\be{\boldsymbol{e}}
\newcommand\oX{\bar{X}}
\newcommand\oZ{\bar{Z}}
\newcommand\obD{\bar{\boldsymbol{D}}}
\newcommand\oD{\bar{D}}
\newcommand\oP{\bar{P}}
\newcommand\obU{\bar{\boldsymbol{U}}}
\newcommand\oU{\bar{U}}
\newcommand\obV{\bar{\boldsymbol{V}}}
\newcommand\oV{\bar{V}}

\newcommand\oSigma{\bar{\Sigma}}
\newcommand\bsigma{\boldsymbol{\sigma}}
\newcommand\bSigma{\boldsymbol{\Sigma}}
\newcommand\obSigma{\bar{\boldsymbol{\Sigma}}}

\newcommand\bI{\mathbf{I}}

\newcommand\bV{\mathbf{V}}

\newcommand\cH{\mathcal{H}}
\newcommand\cK{\mathcal{K}}
\newcommand\cN{\mathcal{N}}
\newcommand{\upd}{\mathrm{d}}

\newcommand\cero{\boldsymbol{0}}

\allowdisplaybreaks 

\begin{document}

\title[{Coupled Stokes~flow with inhomogeneous~poroelasticity}] {Coupling Stokes flow with inhomogeneous poroelasticity}

\author[M. Taffetani]{Matteo Taffetani}
\address{Mathematical Institute, University of Oxford, Radcliffe Observatory Quarter, Woodstock road, OX26GG Oxford, UK. Present address: Department of Engineering Mathematics, University of Bristol, BS8 1TW,Bristol, UK}
\author[R. Ruiz-Baier]{Ricardo Ruiz-Baier}
\address{Mathematical Institute, University of Oxford, Radcliffe Observatory Quarter, Woodstock road, OX26GG Oxford, UK. Present address: School of Mathematical Sciences, Monash University, 9 Rainforest Walk, Melbourne 3800 VIC, Australia}
\author[S. Waters]{Sarah  Waters}
\address{Mathematical Institute, University of Oxford, Radcliffe Observatory Quarter, Woodstock road, OX26GG Oxford, UK}

\date{\today}

\maketitle


\begin{abstract} 
		We investigate the behaviour of flux-driven flow through a single-phase fluid domain coupled to a biphasic poroelastic domain. The fluid domain consists of an incompressible Newtonian viscous fluid while the poroelastic domain consists of a linearly elastic solid filled with the same viscous fluid. The material properties of the poroelastic domain, \textit{i.e.}~permeability and elastic parameters, depend on the inhomogeneous initial porosity field. We identify the dimensionless parameters governing the behaviour of the coupled problem: the ratio between the magnitudes of the driving velocity and the Darcy flows in the poroelastic domain, and the ratio between the viscous pressure scale and the size of the elastic stresses in the poroelastic domain.
		
		We consider a perfusion system, where flow is forced to pass from the single-phase fluid to the biphasic poroelastic domain. We focus on a simplified two dimensional geometry with small aspect ratio, and perform an asymptotic analysis to derive analytical solutions. The slender geometry is divided in four regions, two outer domains that describe the regions away from the interface and two inner domains that are the regions across the interface. Our analysis advances the quantitative understanding of the role of heterogeneous material properties of a poroelastic domain on its mechanical response when coupled with a fluid domain. The analysis reveals that, in the interfacial zone, the fluid and the elastic behaviours of this coupled Stokes - poroelastic problem can be treated separately via (i) a Stokes - Darcy coupling and (ii) the solid skeleton being stress free. This latter finding is crucial to derive the coupling condition across the outer domains for both the elastic part of the poroelastic domain and the fluid flow. Via specification of heterogeneous material properties distribution, we reveal the effects of heterogeneity and deformability on the mechanics of the poroelastic domain.
	\end{abstract}

	\section{Introduction}
	
	In many biological and industrial applications, regions of poroelastic material are surrounded by a single-phase fluid. Examples include the growth of an artificial poroelastic tissue within a perfusion bioreactor \cite{ElHaj1990}, the mechanics of a bacterial biofilm  in a surrounding fluid \cite{Thomen2017}, and  oil/gas extraction from fractured reservoirs \cite{ambar19}.

	In this paper, we provide fundamental insights into the mechanical behaviour of these multiphase/multiregion systems. We focus on a simplified setting where an incompressible Newtonian viscous fluid and a biphasic poroelastic material interact via conditions specified at their interface. The poroelastic domain consists of intrinsically incompressible fluid and solid phases. The permeability and elastic properties depend on the underlying porosity, which is assumed to be spatially varying.
	
	The mechanics of a porous domain is usually described using the theory of mixtures \cite{coussy2004} or the theory of poroelasticity \cite{cowin2007}. These two approaches are equivalent in many contexts, as the one addressed in this paper, but it is possible to highlight some fundamental differences. The theoretical framework underlying the former is suitable in the case of media with many constituents since the primary variables are the volume fraction and the velocity of each constituent. The latter is more straightforward when applied to biphasic media because the domain is seen as a composite material (or effective medium) having a fluid pressure and a solid displacement as primary variables. The pioneering works on biphasic poroelastic media developed by Terzaghi \cite{terzaghi} and Biot \cite{biot41} evaluated the simplest case where the domain undergoes small strains, and thus a linear elastic relation for the effective solid behaviour can be used, and the physical properties (the elastic parameters and the permeability) are constant and isotropic. Later works extended this approach to include kinematic nonlinearity \cite{coussy2004,lang16}, constitutive nonlinearity in the description of the elastic phase and the use of nonlinear and anisotropic permeability field depending on the tortuosity of the porous network and the domain anisotropy \cite{ateshian10,federico07,nedjar13}. There exist several levels of nonlinearity in the description of a poroelastic domain, i.e., kinematic (or geometrical) or constitutive: for a comprehensive discussion we refer  to  \cite{macminn2016}. 
	
	The effect of the heterogeneity has been previously investigated using homogenisation techniques, where the upscaled equations that describe the mechanics of a heterogeneous poroelastic domain are formally derived once the presence of several separated scales is assumed \cite{Mei1997,Saez1989}. Here we state the governing equations at the macroscale, and specify how the material properties depend on porosity via constitutive laws. The interfacial conditions that describe the coupling between a fluid and a poroelastic domain are still a subject of ongoing research \cite{serpilli19,showalter05,murad01} and here we use the conditions proposed in \cite{showalter05}. 
	
	We apply the theoretical framework to a geometrical setup consisting of a two-dimensional thin perfusion system where a fluid flow is forced to pass from a single phase fluid channel into a rectangular poroelastic domain (also known as the consolidation problem). The thin geometry limit allows us to develop a formal asymptotic analysis to obtain closed form solutions (away from the interfacial zone) and reduced formulations (within the interfacial zone) for the mechanics of this coupled system. We extend the work of \cite{Dalwadi2016}, who considered the coupling between a single phase fluid domain and a rigid homogeneous porous domain, to the case of a deformable porous domain with heterogeneous properties. We provide closed form (implicit or explicit) relations between the spatial distribution of the material properties and the displacement, pressure, velocity and stress field; the results from this reduced modelling approach provide mechanistic insights on the behaviour of coupled Stokes flow-poroelastic systems in slender geometries without the need to run full numerical simulations.
	
	We validate the result of the asymptotic analysis using a new mixed finite element method where the poroelasticity equations are written using three fields (displacement, fluid pressure and total pressure).  The unique solvability of the associated continuous and discrete problems, and error estimates, have been recently established in \cite{ruiz21}. With respect to existing approaches, this scheme has two advantages: (i) because of the structure of the weak forms, the imposition of interface conditions does not require Lagrange multipliers across the interface; (ii) the mixed finite element formulation is robust with respect to the Lam\'e first parameter (so the convergence and stability remain valid independently of this parameter). 
	
	The paper is organised as follows. In section~\S\ref{sec:model} we present the equations that govern the poroelastic domain, the fluid domain and the interfacial conditions. We indicate where the non-uniform undeformed porosity enters the governing equations, and highlight the two dimensionless parameters that control the behaviour of this coupled system. In section~\S\ref{sec:channel}, we consider the model in the setting of perfusion in a thin two dimensional geometry. We assume that the aspect ratio (ratio of channel width to length) is small and in section~\S\ref{sec:analytical}  provide closed-form analytical relationships that enables characterisation of the system behaviour without solving the full fluid-poroelastic coupled problem numerically. In section~\S\ref{sec:result} we select a set of geometrical and constitutive parameters to investigate the analytical solution in the steady and oscillatory cases, which we validate via comparison with finite element simulations. We close with some remarks and possible extensions in section~\S\ref{sec:concl}.

	\section{Governing equations}\label{sec:model}

\begin{figure}[t]
		\begin{center}
			\includegraphics[height=0.35\textwidth]{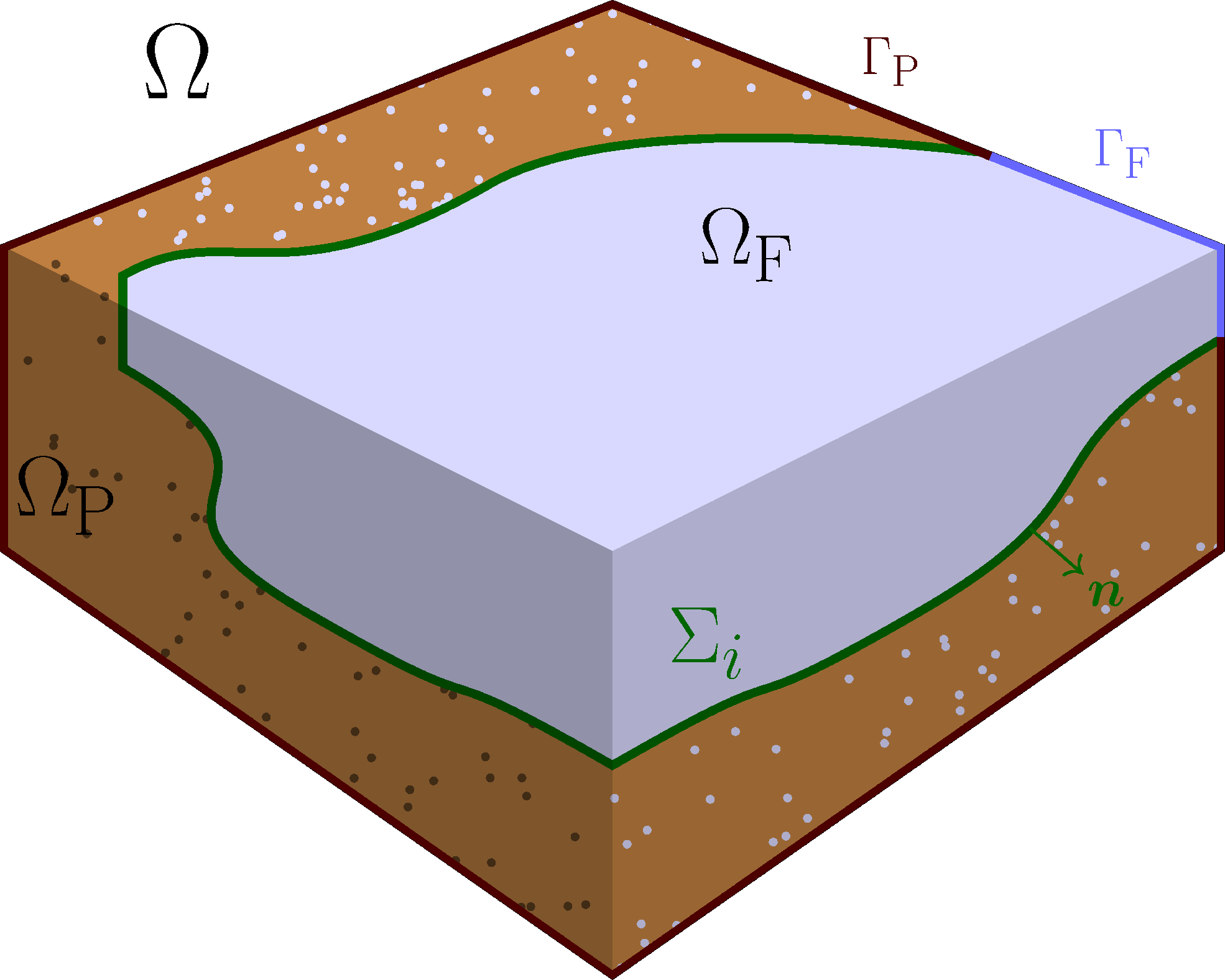}
		\end{center}
		\caption{Schematic diagram of the multidomain configuration. The blue region is the fluid domain $\Omega_F$; the brown region is the poroelastic domain $\Omega_P$; the green curve is the interface $\Sigma_i$. The blue boundary represents $\Gamma_F$ and the purple boundary represents $\Gamma_P$.}\label{fig:sketch} 
	\end{figure}

	With reference to fig.~\ref{fig:sketch}, we consider a  domain $\Omega\subset \mathbb{R}^d$, $d=2,3$, divided into subdomains $\Omega_F$ and $\Omega_P$ representing incompressible, viscous Newtonian fluid and poroelastic solid regions, respectively. The subdomain boundaries are $\partial\Omega_F$ and $\partial\Omega_P$, respectively. The interface between the two subdomains is  $\Sigma_i=\Omega_F\cap \Omega_P$ and  the unit normal vector to the interface is $\nn$ (with the convention that it points from $\Omega_F$ to $\Omega_P$). We denote the external boundaries of the subregions by $\Gamma_F = \partial\Omega_F\setminus\Sigma_i$ and 
	$\Gamma_P = \partial\Omega_P \setminus\partial\Sigma_i$.  Spatial coordinates are denoted by $\bx$ while $t$ denotes time.

	We assume the fluid and solid constituents are intrinsically incompressible, i.e., the true densities are constant. We neglect gravitational and inertial effects throughout. 
	The governing equations and interfacial conditions are presented in a dimensionless form motivated as follows. Throughout the system, we have an incompressible, viscous, Newtonian fluid with dynamic viscosity $\mu_f$. The poroelastic domain has permeability
	$\hat{\kappa}$, Lam\'e parameters of size $\hat{\mu}_s$, and characteristic displacements of size $\delta$. Assuming the spatial domain to have characteristic length $L$, the stresses in the poroelastic domain are of size $\hat{\mu}_s\delta/L$, and the size of the Darcy flows generated by gradients in stress are then $\hat{\kappa}\delta \hat{\mu}_s/(\mu_f L^2)$. The characteristic time $\tau$ is chosen to ensure the  Darcy and solid skeleton velocities are comparable, i.e., that $\delta/\tau\sim\hat{\kappa}\delta \hat{\mu}_s/(\mu_f L^2)$, and so $\tau\sim (\mu_f L^2)/(\hat{\kappa}\hat{\mu}_s)$. Assuming flows of size $U_{av}$ in the single-phase domain, the viscous pressure scale in given by $U_{av}\mu_f/L$. 
	
	\subsection{Poroelastic domain}\label{poro}
	The governing equations for a saturated linearly elastic porous domain can be written as a system of two equations in two unknowns, the fluid pressure $p_P\left(\bx,t\right)$ and the solid phase displacement $\bd\left(\bx,t\right)$. This system reads \cite{coussy2004,Preziosi2002,macminn2016}
	\begin{subequations}
		\label{short}
		\begin{align}
		\label{eq:balanceDIM}
		\nabla \cdot \left(\bsigma'- p_P \bI\right) &= \cero,\\
		\label{eq:continuityDIM}
		\nabla\cdot\left(\frac{\partial \bd}{\partial T} +\bv_d\right) & =0.
		\end{align}
	\end{subequations}
	In equation \eqref{eq:balanceDIM}, $\bsigma'$ is the Terzaghi stress tensor given by
	\begin{equation}\label{eq:linear_elasticity_dimensional}
	\bsigma'\left(\bx,t\right) = 2 \mu_s (\bx,t) \be(\bd) + \lambda(\bx,t) \left(\nabla \cdot \bd\right) \bI, 
	\end{equation}
	where $\be(\bd) = (\nabla\bd+\nabla\bd^{\tt t})/2$ is the 
	infinitesimal strain tensor and $\lambda,\mu_s$ are the Lam\'e constants of the solid, which we assume  vary with the initial porosity field $\theta_0(\bx)$ via 
	\begin{align}
	\mu_s = \hat{\mu}_s\cH_0\left(\theta_0\right), \quad \lambda = \hat{\mu}_s\cN_0\left(\theta_0\right),
	\end{align}
	where $\hat{\mu}_s$ is constant, and $\cH_0(\theta_0)$ and $\cN_0(\theta_0)$ are $O(1)$ functions.
	In equation \eqref{eq:continuityDIM} the Darcy velocity $\bv_d\left(\bx,t\right)$ is defined as
	\begin{equation}\label{eq:vd}
	\bv_d = - \frac{\hat{\kappa} \cK_0}{\mu_f}\nabla p_P,
	\end{equation}
	where we rewrite the permeability as $\kappa(\theta_0) = \hat{\kappa} \cK_0(\theta_0)$ with $\hat{\kappa}$ that captures the size of the permeability and $\cK$ being an $O(1)$ function reflecting the dependence of the permeability on the spatially varying porosity field. Motivated by the discussion in \S\ref{sec:model}, we
	use
	\begin{equation}\label{EQ:dimensionless_variables}
	\bx = L \bX, \quad \bd = \delta \bD, \ t=\frac{\mu_f L^2}{\hat{\kappa}\hat{\mu}_s} T, \ \left[\bsigma',p_p\right] = \frac{\hat{\mu}_s\delta}{L} \left[\bSigma',P_P\right], \ \bv_d=\frac{\hat{\kappa}\delta\hat{\mu}_s}{\mu_f L^2}\bV,
	\end{equation}
	to obtain the following system of dimensionless equations
	\begin{subequations}\label{eq:duo1}
		\begin{align}
		\label{EQ:balance_final}
		\nabla \cdot \left(\bSigma' - P_P \bI\right) &= \cero, \\ 
		\label{eq:vd_dimensionless}
		\bV  &= - \cK_0\left(\mathbf{X}\right)\nabla P_P, \\
		\label{EQ:continuity_final}
		\nabla\cdot\left(\frac{\partial \bD}{\partial T} +\bV\right) & =0,  \\
		\bSigma' & = 2 \cH_0\left(\bX\right) \be(\bD) + \cN_0\left(\bX\right) (\nabla \cdot \bD) \bI, 
		\end{align}
	\end{subequations}
	where $\be({\bD})=(\nabla \bD+\nabla \bD^{\rm{t}})/2$. The functions $\cK_0$, $\cH_0$ and $\cN_0$ depend on the undeformed porosity $\theta_0$ which, in turn, is a function of $\bX$. For this reason, in the remainder of the paper, we write these functions directly with respect to the spatial coordinates.
	We emphasise that the variation of porosity $\theta_1(\bX,T)$ can, in the linear theory, be computed a posteriori once the displacement field is known. Consequently, such 
	variation can be written as 
	\begin{equation}\label{eq:theta1} 
	\theta_1 = \nabla \cdot \left(\left(1-\theta_0\right) \bD_0\right).
	\end{equation}
	Equation \eqref{eq:theta1} reveals how the pore structure is rearranged in response to deformation of the poroelastic material. 
	As an example, let us consider a solenoidal displacement field $\nabla \cdot \bD = 0$: in the case of a homogeneous initial porosity we necessarily have $\theta_1 = 0$, thus the porosity field is unaffected; in the case of heterogeneous initial porosity, instead, this displacement causes a rearrangement of the pores of the form $\theta_1 = -\nabla \theta_0 \cdot \bD$. 
	
	\subsection{Fluid domain}\label{fluid}
	In the single-phase domain, the fluid is governed by the  Stokes and continuity equations. We let $U_{av}$ denote the velocity scale, and non-dimensionalise the  velocity via $\bu=U_{av}\bU$ and pressure via $p_f=(\mu_f U_{av}/L) P_F$. The dimensionless  governing equations are then 
	\begin{subequations}\label{EQ:duoF}
		\begin{align}
		\label{eq:momentumA_dimensionless}
		-\nabla \cdot[2\be(\bU) - P_F\bI] & = \cero, \\ 
		\label{eq:massA_dimensionless}
		\nabla \cdot \bU & = 0.
		\end{align}\end{subequations}
	\subsection{Boundary and interfacial conditions}\label{int}
	To couple the equations in the poroelastic \eqref{eq:duo1} and  fluid \eqref{EQ:duoF} domains, we impose boundary and interfacial conditions. The boundary conditions are specific to the physical problem under consideration and we delay their specification until \S \ref{sec:channel}. We specify the following interfacial conditions (see \cite{showalter05}): the requirements of continuity of traction and that the normal component of the stress in the fluid domain is equal to the pressure
	in the poroelastic material, together with the Beavers-Joseph-Saffman condition, imply that 
	\[\nn\cdot\bsigma'\nn = 0,\qquad -\bt\cdot(2\mu_f \be(\bu))\nn = \frac{\gamma \mu_f}{\sqrt{\hat{\kappa}\cK}} \left(\bu - \frac{\partial \bd}{\partial t} \right)\cdot \bt,\qquad p_F - p_P = 2\mu_f \nn\cdot \be(\bu)\cdot \nn,\]
	and mass conservation yields 
	\[\bu\cdot \nn=\left(\frac{\partial \bd}{\partial t}  +\bv_d\right)\cdot \nn,\]
	where $\gamma>0$ is the slip coefficient, and we recall that the normal $\nn$ on the interface is understood as pointing from the fluid domain $\Omega_F$ towards the porous structure $\Omega_P$, while $\bt$ denotes the 
	tangent vector(s) to $\Sigma_i$ (one tangent for the case of $d=2$, and two tangent vectors  normal to $\nn$ for $d=3$). In dimensional form, these conditions can be rewritten as
	\begin{subequations}
		\begin{align}
		\label{eq:inter-U}
		&\bu\cdot\nn  = \left(\frac{\partial \bd}{\partial t}  - \frac{\hat{\kappa} \cK}{\mu_f} \nabla p_P \right)\cdot \nn,\\
		\label{eq:inter-sigma}
		&(2\mu_f \be(\bu) - p_F\bI)\cdot \nn  = (\bsigma'- p_P \bI ) \cdot  \nn,\\
		\label{eq:inter-sigmaf}
		&-\nn\cdot \left[(2\mu_f \be(\bu) - p_F\bI)\cdot\nn\right]  =  p_P,\\
		\label{eq:inter-bjs}
		&- \nn\cdot\left[ (2\mu_f \be(\bu) - p_F\bI)\cdot\bt \right] = \frac{\gamma\mu_f}{\sqrt{\hat{\kappa}\cK}} \left(\bu - \frac{\partial \bd}{\partial t} \right)\cdot \bt.
		\end{align}\end{subequations}

	Non-dimensionalising as in sections \S\ref{poro} and \S\ref{fluid}, the interface conditions are
	\begin{subequations}
		\label{eq:interf00}
		\begin{align}
		\label{eq:inter-U_dimensionless_v1}
		&U_n \bU\cdot\nn  = \left(\frac{\partial \bD}{\partial T} - \cK_0 \nabla P_P \right)\cdot \nn,  & \\
		\label{eq:inter-sigma_dimensionless_v1}
		&U_nK_n(2 \be(\bU) -  P_F\bI)\cdot \nn = (\bSigma' - P_P \bI ) \cdot \nn,  & \\
		\label{eq:inter-sigmaf_dimensionless_v1}
		&-U_nK_n\nn\cdot (2  \be(\bU) -  P_F\bI)\cdot\nn  =  P_P, & \\
		\label{eq:inter-bjs_dimensionless_v1}
		&- U_n\nn\cdot (2 \be(\bU) - P_F\bI)\cdot\bt = \frac{\gamma}{\sqrt{K_n \cK_0}} \left(U_n \bU - \frac{\partial \bD}{\partial T}\right)\cdot \bt,  & 
		\end{align}\end{subequations}
	where $K_n=\hat{\kappa}/L^2$ is the ratio between the area occupied by the pores and the characteristic cross section $L^2$ of the poroelastic domain \cite{Dalwadi2016}, $U_n=U_{av}\mu_f L^2/(\delta\hat{\mu}_s\hat{\kappa})$ is the ratio of velocity scales in the Stokes and poroelastic domains,  and $K_n U_n$ captures the ratio between pressure scales in the Stokes and Darcy domains. 
	
	\section{Perfusion in a thin two-dimensional geometry}\label{sec:channel}
	We apply the framework from \S \ref{sec:model} to a two-dimensional channel where the flow is forced to pass through a single-phase fluid region into a poroelastic domain. 
	
	\begin{figure}[t!]
		\centering
		\includegraphics[width=0.99\textwidth]{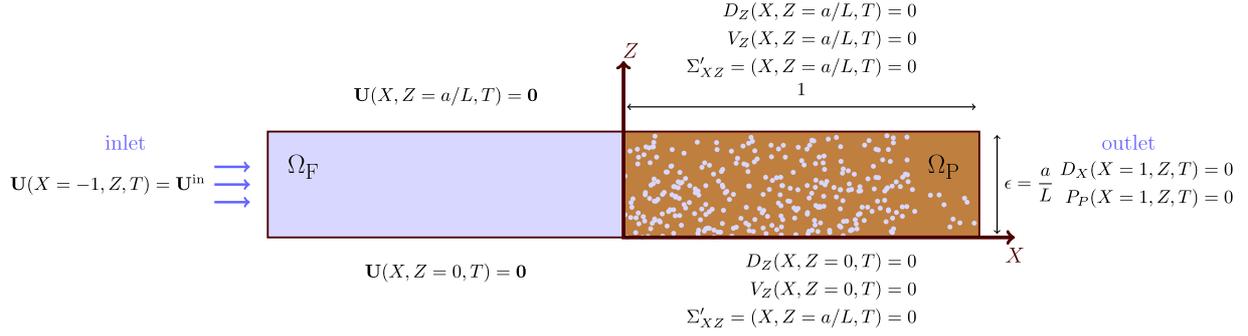}
		\caption{Sketch of the two-dimensional channel geometry in rescaled variables, indicating the single-phase fluid and poroelastic regions, together with the boundary conditions.}
		\label{fig:channel}
	\end{figure}
	
	The problem setup (in dimensionless variables) is illustrated in fig. \ref{fig:channel}.   We consider a thin-channel, with width $a$ small compared to the characteristic length $L$.  We drive the flow via the specification of an upstream time-dependent velocity, and prescribe zero pressure and zero axial displacement at the outlet.  We adopt a 2D Cartesian coordinate system $\bX=(X,Z)$, where lengths are non-dimensionalised with respect to $L$. The corresponding coordinate directions are given by $\mathbf{e}_x$ and $\mathbf{e}_z$ respectively.  We denote components of vectors via $\bU = \left(U_{X},U_{Z}\right)^{\tt t}$, $\bD = \left(D_X,D_Z\right)^{\tt t}$, $\bV = \left(V_{X},V_{Z}\right)^{\tt t}$ and components of the stress tensor $\bSigma'$ are denoted $\Sigma'_{XX}, \Sigma'_{XZ}$ and $\Sigma'_{ZZ}$. 
	
	The governing equations (\ref{eq:duo1}) and (\ref{EQ:duoF}) are coupled via the interfacial conditions (\ref{eq:interf00}). Additionally, we specify the following boundary conditions. For the single-phase fluid no-slip boundary conditions are applied at the side walls $Z=0,a/L$.  In the poroelastic domain we impose zero transverse displacement, zero normal Darcy flow, and zero elastic shear stress at $Z=0,a/L$.  We prescribe a parabolic axial velocity profile $\bU_{in}(X=-1,Z,T)=-6 (U_{in}/U_{av})f(T)(Z^2-Z)\mathbf{e}_x$ at the inlet $X=-1$, with $f(T)$ describing the time dependence of the imposed velocity profile, and $(U_{in}/U_{av})f(T)$ is the dimensionless inlet flux; in the steady case the time dependence reduces to $f(T)=1$ while, in the oscillatory one, one can set $f(\Omega T)$ with $\Omega=\omega \left(\mu_f L^2/\left(\hat{\kappa}\hat{\mu}_s\right)\right)$ being the dimensionless pulsation (and $\omega$ the dimensional counterpart) of the forcing term. At the outlet $X=1$, we prescribe $P_P(X=1,Z,T)=0$ and $D_X (X=1,Z,T)=0$, allowing the fluid to flow away from the poroelastic domain but the solid phase to be constrained. The undeformed interface is  located at $X=0$ and the unit normal and tangent vectors are $\nn = \mathbf{e}_x$ and $\bt = \mathbf{e}_z$.

	We exploit that the channel is thin, i.e., $a \ll L$, and introduce the small parameter $\epsilon = a/L$.  The system is then characterised by the dimensionless parameters $\epsilon$, $K_n$, $U_n$ and $U_{in}/U_{av}$. Rewriting $K_n=\epsilon^2 (\kappa/a^2)$ and requiring that $\bar{K}=\kappa/a^2$ is (at most) $O(1)$, we see that $K_n=\epsilon ^2 \bar{K}$. Requiring the pressures in the single-phase fluid and poroelastic domains to be of the same size we see that $U_n K_n=O(1)$, and we set $U_n=\epsilon^{-2}$.  To be consistent with the lubrication scalings given below in (\ref{lub}), we additionally choose $U_{in}/U_{av}=\epsilon^2$.

	\section{Analytical solution} \label{sec:analytical}
\begin{figure}[!t]
		\centering
		\includegraphics[width=0.99\textwidth]{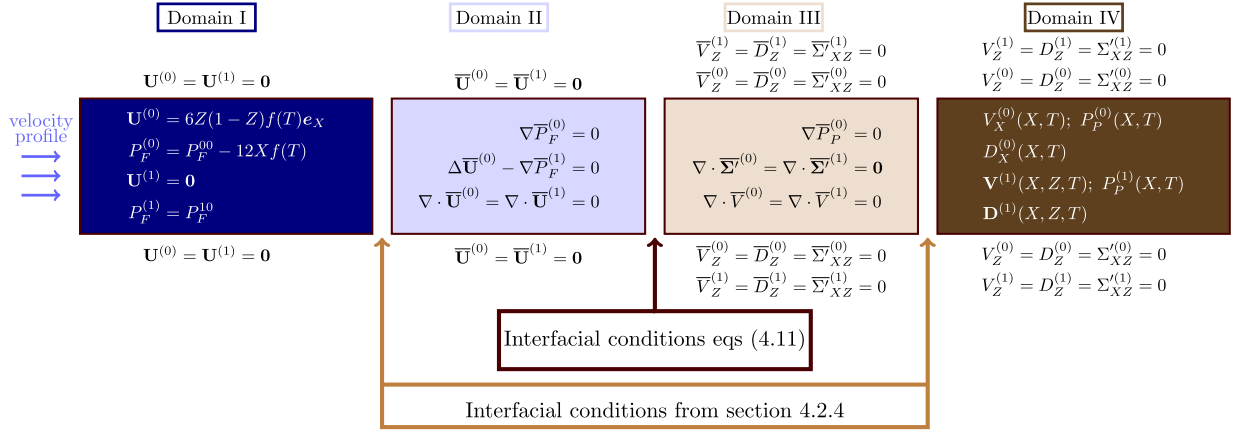}
		\caption{Sketch of asymptotic structure. The flow goes from left to right. Boxes I and II represent the fluid domain (outer - inner) while boxes III and IV are the poroelastic domain (inner - outer). The outer domains reduce to a Poiseuille - poroelastic system coupled through the interface conditions from \S\ref{SECSEC:Transmission}. In the lubrication limit, the inner domains reduce to be a Stokes - rigid Darcy system with interface conditions derived in equations \eqref{eq:inter-U_lubrication_inner}-\eqref{eq:inter-bjs__lubrication_inner} decoupled with a divergence free stress field. Solution structures/governing equations are shown in the boxes.}
		\label{FIG:MoStructure}
	\end{figure}

	We start from the equations (\ref{eq:duo1}) and (\ref{EQ:duoF}) in the lubrication regime when the axial lengthscale ($O(1)$) is large compared with the transverse ($O(\epsilon)$) lengthscale. This assumption breaks down in the inner regions either side of the interface (where the axial and transverse lengthscales are  comparable).
	Following \cite{Dalwadi2016}, we anticipate the presence of four regions, two outer domains of (dimensionless) length 1 (one fluid (domain I) and one poroelastic (domain IV)) and two inner domains of dimensionless length $\epsilon$ (one fluid (domain II) and one poroelastic (domain III)) as highlighted in fig.~\ref{FIG:MoStructure}.  The interface conditions \eqref{eq:interf00} are applied in the inner domain.  We solve the system in the inner domains (one fluid (domain II) and one poroelastic (domain III)), consider their far field behaviour and match with the outer domain solutions, to derive conditions coupling the outer problems together. The asymptotic analysis is carried out to order $O(\epsilon)$ for two reasons: (\textit{i}) the effect of porosity heterogeneities in the transverse direction appear at the order $O(\epsilon)$; and (\textit{ii}) a complete description of the leading-order behaviour in the inner regions requires  information at $O(\epsilon)$ in the outer regions.  Below we consider each domain separately.

	\subsection{Asymptotic structure}
	
	\subsubsection{Outer domains I and IV}\label{SEC:Outer}
	In the outer domains, the axial length scale is large compared to the transverse lengthscale. We rescale as follows
	\begin{equation}\label{lub}
	\begin{gathered}
	(X,Z)=(\tilde{X},\epsilon \tilde{Z}), \ (D_X,D_Z) = (\tilde{D}_X,\epsilon \tilde{D}_Z),\ (U_X, U_Z)=(\epsilon^2\tilde{U}_X,\epsilon^3 \tilde{U}_{Z}), \nonumber \\
	(V_X, V_Z)=(\tilde{V}_X, \epsilon\tilde{V}_Z), (P_F, P_P)=(\tilde{P}_F, \tilde{P}_P), \ (\Sigma'_{XX}, \Sigma'_{XZ}, \Sigma'_{ZZ})=(\tilde{\Sigma}'_{XX}, \epsilon^{-1}\tilde{\Sigma}'_{XZ}, \tilde{\Sigma}'_{ZZ}). 
	\end{gathered}
	\end{equation}
	
	Dropping the tildes, \eqref{EQ:balance_final}, \eqref{eq:vd_dimensionless} and \eqref{EQ:continuity_final} in the poroelastic domain are now
	\begin{subequations}\label{eq:taoT}
		\begin{align}
		\label{EQ:equilibrium_LL}
		\epsilon^2\frac{\partial \Sigma_{XX}'}{\partial X} + \frac{\partial \Sigma_{XZ}'}{\partial Z} - \epsilon^2 \frac{\partial P_P}{\partial X}= 0, \quad & \frac{\partial \Sigma_{XZ}'}{\partial X} + \frac{\partial \Sigma_{ZZ}'}{\partial Z} -\frac{\partial P_P}{\partial Z} = 0,&\\ 
		\label{EQ:VD_LL}
		V_X = - \cK_0 \frac{\partial P_P}{\partial X}, \quad & \epsilon^2 V_Z = - \cK_0 \frac{\partial P_P}{\partial Z}, & \\
		\label{EQ:continuity_LL}
		\frac{\partial}{\partial X} \left(\frac{\partial D_X} {\partial T} + V_X\right) &+\frac{\partial}{\partial Z} \left(\frac{\partial D_Z} {\partial T} + V_Z\right) =0,& 
		\end{align}\end{subequations}
	with the entries of the symmetric stress tensor $\bSigma'$ specified as
	\begin{subequations}\label{EQ:stress_outer}
		\begin{align}
		\Sigma'_{XX} &= 2 \cH_0 \frac{\partial D_X}{\partial X} + \cN_0 \left( \frac{\partial D_X}{\partial X} +  \frac{\partial D_Z}{\partial Z} \right), \\
		\Sigma'_{ZZ} &= 2 \cH_0 \frac{\partial D_Z}{\partial Z} + \cN_0 \left( \frac{\partial D_X}{\partial X} +  \frac{\partial D_Z}{\partial Z} \right),\\
		\Sigma'_{XZ} &= \cH_0 \left(\frac{\partial D_X}{\partial Z} + \epsilon^2  \frac{\partial D_Z}{\partial X} \right).
		\end{align}
	\end{subequations} 
	
	The rescaled Stokes equations \eqref{eq:momentumA_dimensionless} and \eqref{eq:massA_dimensionless} are
	\begin{subequations}\label{eq:taoM}
		\begin{align}
		\label{EQ:equilibriumF_LL}
		\epsilon^2 \frac{\partial^2 U_X}{\partial X^2} + \frac{\partial^2 U_X}{\partial Z^2} - \frac{\partial P_F}{\partial X} = 0, \  & \epsilon^4 \frac{\partial^2 U_Z}{\partial X^2} + \epsilon^2 \frac{\partial^2 U_Z}{\partial Z^2}- \frac{\partial P_F}{\partial Z} = 0, & \\ 
		\label{EQ:continuityF_LL}
		\frac{\partial U_X}{\partial X} &+ \frac{\partial U_Z}{\partial Z} = 0. & 
		\end{align}\end{subequations}

	The boundary conditions are 
	\begin{subequations}
		\label{plant}
		\begin{align}
		\label{EQ:BC_noslipF}
		(U_X, U_Z)=(0,0) \quad \text{on} \quad Z=0,1 & \quad \text{for} \quad X<0, \\ 
		\label{EQ:BC_noslipP}
		(D_Z, V_{Z},\Sigma_{XZ}')=(0,0,0)  \quad \text{on} \quad Z=0,1 & \quad \text{for} \quad X>0,\\
		\label{EQ:BC_outlet}
		D_X = 0,\quad P_P = 0 & \quad \text{at} \quad X = 1,\\
		\label{EQ:BC_inlet}
		(U_{X}, U_{Z}) =(-6f(T)(Z^2-Z),0) & \quad \text{at} \quad X =-1.
		\end{align}\end{subequations}
	
	We pose an asymptotic expansion, as $\epsilon\rightarrow 0$,  of the form
	\begin{align}
	\label{EQ:expansion_LL}
	\nonumber \left(\bU, P_F, \bD, P_P, \bV,\Sigma_{IJ}\right) & = \left(\bU^{(0)}, P_F^{(0)}, \bD^{(0)}, P_P^{(0)}, \bV^{(0)}, \Sigma_{IJ}^{(0)}\right)\\ & \qquad +\epsilon\left(\bU^{(1)}, P_F^{(1)}, \bD^{(1)}, P_P^{(1)}, \bV^{(1)}, \Sigma_{IJ}^{(1)}\right) + O(\epsilon^2),
	\end{align}
	where $I,J\in{X,Z}$.
	The function $\cH_0 \left(X, \epsilon Z\right)$ is expanded as
	\begin{align}
	\label{EQ:expansionH_LL}
	\cH_0 \left(X, \epsilon Z\right) &= \cH_0^{(0)} \left(X\right) +\epsilon\left( \frac{\partial \cH_0 }{\partial \epsilon}\big\vert_{\epsilon=0}\right)+ O(\epsilon^2)= \cH_0^{(0)} \left(X\right) +\epsilon \cH_0^{(1)} \left(X, Z\right)+ O(\epsilon^2),
	\end{align}
	and for $\cK_0 \left(X, \epsilon Z\right)$ and $\cN_0 \left(X, \epsilon Z\right)$ we proceed analogously. We note that the leading-order fields are functions of $X$ only.
	
	\subsubsection{Inner domains II and III}
	In the inner domains,  the axial and transverse lengthscales are comparable, and we rescale as follows
	\begin{equation}
	\begin{gathered}
	(X,Z)=\epsilon(\bar{X},\bar{Z}), \ (D_X,D_Z) = (\bar{D}_X, \bar{D}_Z),\ (U_X, U_Z)=(\epsilon^2\bar{U}_X,\epsilon^2 \bar{U}_{Z}), \nonumber \\
	(V_X, V_Z)=(\bar{V}_X, \bar{V}_Z),\  (\Sigma'_{XX}, \Sigma'_{XZ}, \Sigma'_{ZZ})=\epsilon^{-1}(\bar{\Sigma}'_{XX}, \bar{\Sigma}'_{XZ}, \bar{\Sigma}'_{ZZ}).
	\end{gathered}
	\end{equation}

	We substitute these rescalings into (\ref{eq:duo1}) and (\ref{EQ:duoF}), and define $\bar{\nabla}=(\partial/\partial \bar{X}, \partial/\partial \bar{Z})$.
	
	In the inner fluid domain II we recover the scenario analysed in \cite{Dalwadi2016}, and thus 
	\begin{equation}\label{EQ:inner_fluid}
	- 2 \epsilon \bar{\nabla} \cdot \be(\obU) - \bar{\nabla} \bar{P}_F  = \cero, \qquad 
	\nabla \cdot \obU  = 0, 
	\end{equation}
	with $\be(\obU)  = \left(\bar{\nabla} \obU + \bar{\nabla} \obU^t\right)/2$. In the inner porous domain III we have
	\begin{equation}\label{EQ:inner_porous}
	\bar{\nabla} \cdot  \left(\obSigma' - \epsilon \oP_P \bI\right) = \cero, \quad
	\epsilon \obV = - \cK_0\bar{\nabla} \oP_P, \quad
	\bar{\nabla} \cdot \left(\frac{\partial \obD}{\partial T} + \obV\right)  =0,
	\end{equation}
	where the components of the symmetric solid stress in the porous domain $\obSigma'(\bar{\bX}, T)$ are
	\begin{subequations}\label{EQ:StressesInner}
		\begin{align}
		\label{EQ:StressesInner1}
		\oSigma_{XX}' &= \left(2\cH_0 + \cN_0\right)\frac{\partial \oD_X}{\partial \oX} + \cN_0 \frac{\partial \oD_Z}{\partial \oZ},\\
		\label{EQ:StressesInner2}
		\oSigma_{ZZ}' &= \left(2\cH_0 + \cN_0 \right)\frac{\partial \oD_Z}{\partial \oZ} + \cN_0 \frac{\partial \oD_X}{\partial \oX},\\
		\label{EQ:StressesInner3}
		\oSigma_{XZ}' &= \cH_0 \left(\frac{\partial \oD_Z}{\partial \oX} + \frac{\partial \oD_X}{\partial \oZ}\right).
		\end{align} 
	\end{subequations}
	
	At the top and the bottom  boundaries, we have the boundary conditions
	\begin{subequations}\label{EQ:BCS_inner}
		\begin{align}
		\obU = \cero \quad \text{on} \quad \oZ=0,1 \quad  & \text{for} \quad \oX<0, \\
		\label{EQ:BCS_innerb}
		(\oD_Z, \oV_{Z}, \oSigma'_{XZ})=(0,0,0)  \quad \text{on} \quad \oZ=0,1 \quad & \text{for} \quad \oX>0.
		\end{align}\end{subequations}
	The interface conditions \eqref{eq:inter-U_dimensionless_v1}--\eqref{eq:inter-bjs_dimensionless_v1} rescale as
	\begin{subequations}\label{EQ:inter-lubrication_inner}
		\begin{align}
		\label{eq:inter-U_lubrication_inner}
		\epsilon \oU_{X} & = \epsilon \frac{\partial \oD_X}{\partial T}  -  \cK_0 \frac{\partial \oP_P}{\partial \oX}, & \\
		\label{eq:inter-sigma_lubrication_inner}
		2 \epsilon^2\bar{K} \frac{\partial \oU_X}{\partial \oX} - \epsilon\bar{K} \oP_F = \oSigma_{XX}' - &\epsilon \oP_P;  \epsilon^2\bar{K} \frac{\partial \oU_X}{\partial \oZ} + \epsilon^2\bar{K} \frac{\partial \oU_Z}{\partial \oX} = \oSigma_{XZ}', & \\
		\label{eq:inter-sigmaf__lubrication_inner}
		- 2 \epsilon\bar{K} \frac{\partial \oU_{X}}{\partial \oX} + \bar{K}\oP_F & =  \oP_P, & \\
		\label{eq:inter-bjs__lubrication_inner}
		\frac{\partial \oU_{X}}{\partial \oZ} +\frac{\partial \oU_{Z}}{\partial \oX} & = \gamma \frac{1}{\sqrt{\bar{K}\cK_0}} \left(\oU_Z - \frac{\partial \oD_Z}{\partial T}\right).  & 
		\end{align}\end{subequations}
	
	In the inner domains, we pose an asymptotic expansion of the form
	\begin{equation*}
	\left(\obU, \oP_F, \obD, \oP_P, \obV\right) = \left(\obU^{(0)}\!\!, \oP_F^{(0)}, \obD^{(0)}, \oP_P^{(0)}, \obV^{(0)}\right)+\epsilon\left(\obU^{(1)}\!\!, \oP_F^{(1)}\!\!, \obD^{(1)}, \oP_P^{(1)}, \obV^{(1)}\right) + O(\epsilon^2),
	\end{equation*}
	as $\epsilon\rightarrow 0$, and, consistently, the function $\cH_0 \left(\oX, \oZ\right)$ is expanded as
	\begin{align}
	\cH_0 \left(\oX, \oZ\right) &= \cH_0^{(0)} \left(0\right) +\epsilon \left(  \frac{\partial \cH_0^{(0)} \left( \epsilon \oX\right)}{\partial \epsilon} \big \vert_{\epsilon=0} + \cH_0^{(1)} \left(0, \oZ\right)\right)+ O(\epsilon^2),
	\end{align}
	and analogously for $\cK_0 \left(\oX, \oZ\right)$ and $\cN_0 \left(\oX, \oZ\right) $.
	Interestingly, the leading-order term of the material properties is constant in the inner porous domain. 
	
	In the following section, we present the asymptotic solutions for the pressure, velocity and deformation fields in all regions. For ease of presentation, we state the derived solutions for domains I and IV in sections~\S\ref{sec:DomainI} and~\S\ref{sec:DomainIV}, and give the full details of the solution derivation in Appendix \ref{APP:A} and \ref{APP:B}. We give full details of the derivation of the solutions in domains II and III in section~\S\ref{sec:DomainsIIIII} . Finally we show how coupling conditions can be derived from the inner solutions in regions II and III that match to the outer solutions in domains I and IV. 
	\subsection{Asymptotic solution}

	\subsubsection{Fluid domain I}\label{sec:DomainI}
	The behaviour in the outer fluid domain is analogous to the case studied in \cite{Dalwadi2016}. The leading and first order solutions in domain I are (see  Appendix \ref{APP:A} for details)
	\begin{subequations}
		\begin{align}
		\label{EQ:Fluid_Outer0}
		&\bU^{(0)}= (6 Z (1-Z) f(T),0), &P_F^{(0)} = P_F^{00} - 12 X f(T),\tag{\theequation a,b}\\
		\label{EQ:Fluid_Outer1}
		&\bU^{(1)}= (0,0), &P_F^{(1)}= P_F^{10}(T).\tag{\theequation c,d}
		\end{align}\end{subequations}
	The pressures contain two unknowns $P_F^{00}$ and $P_F^{10}(T)$ to be determined via matching the leading and first order pressure fields to the solutions in the inner domain (region II). 
	
	\subsubsection{Poroelastic domain IV}\label{sec:DomainIV}
	The leading-order solutions in domain IV are (see Appendix \ref{APP:B})
	\begin{subequations}
		\begin{align}
		\label{EQ:PorousD_Outer0}
		&\bD^{(0)} = (A^{(0)}_D(X,T),0), \\
		\label{EQ:PorousP_Outer0}
		&P_P^{(0)} = P_P^{01}(T)\int_1^X \frac{1}{\cK_0^{(0)}}\upd \xi + \int_1^X  \frac{1}{\cK_0^{(0)}}\left(\int_{0}^{\xi}\left(\frac{\partial^2}{\partial T \partial \eta}A^{(0)}_D(\eta,T)\right)\upd \eta\right)\upd \xi,\\
		\label{EQ:PorousV_Outer0}
		&\bV^{(0)} =\left( - P_P^{01}(T) + \int_0^X \frac{\partial^2}{\partial T \partial \eta}A^{(0)}_D(\xi,T)\upd \xi,0\right).
		\end{align}\end{subequations}
	
	We see that the leading-order fields are determined up to the specification of the functions $A_D^{(0)}(X,T)$, that depends on both the longitudinal coordinate $X$ and the time $T$, and $P_P^{01}(T)$, that depends only on time $T$.
	
	The first order solutions in domain IV are (see Appendix \ref{APP:B})
	\begin{subequations}
		\begin{align}
		\label{EQ:PorousD_Outer1}
		&\bD^{(1)} = \left(A^{(1)}_D(X,T),\frac{1}{2 \cH_0^{(0)}+\cN_0^{(0)}}\left( Z\int_0^1\cN_0^{(1)}\upd \zeta-\int_0^Z \cN_0^{(1)}\upd \zeta\right)\frac{\partial A_D^{(0)}}{\partial X}\right), \\
		\label{EQ:PorousP_Outer1}
		&P_P^{(1)} = P_P^{11}(T)\int_1^X \frac{1}{\cK_0^{(0)}}\upd \xi + \int_1^X  \frac{1}{\cK_0^{(0)}}\left(\int_0^{\xi}\left(\frac{\partial^2 A_D^{(1)}}{\partial T \partial X}  - \frac{\cK_0^{(1)}}{\cK_0^{(0)}}\frac{\partial^2 A_D^{(0)}}{\partial T \partial X} \right.\right. \nonumber\\
		&\left.\left.- \frac{1}{\cK_0^{(0)}}\int_0^1\cK_0^{(1)}\upd \zeta\left(P_P^{01} + \int_0^X \frac{\partial^2A_D^{(0)}}{\partial T \partial \xi} \upd \xi\right)\right)\upd \eta\right)\upd \xi,\\
		&\bV^{(1)} =\left( -  \left(\cK_0^{(0)} \frac{\partial P_P^{(1)}}{\partial X}   + \cK_0^{(1)} \frac{\partial P_P^{(0)}}{\partial X}  \right),\right.\nonumber\\
		\label{EQ:PorousV_Outer1}
		&\left.-Z \frac{\partial^2 A_D^{(1)}}{\partial T \partial X} + \int_0^Z \frac{\partial}{\partial X}  \left(\cK_0^{(1)} \frac{\partial P_P^{(0)}}{\partial X}\right) \upd \zeta + Z \frac{\partial}{\partial X}  \left(\cK_0^{(0)} \frac{\partial P_P^{(1)}}{\partial X}\right)\right).
		\end{align}\end{subequations}
	where $\bV^{(1)}$ is left in terms of $P_P^{(0)}$ and $P_P^{(1)}$ for the sake of conciseness.
	
	The first order fields are specified up to the unknown functions $A_D^{(0)}(X,T)$ and $P_P^{01}(T)$, already introduced at the leading-order, and the functions $A_D^{(1)}(X,T)$ and $P_P^{11}(T)$ that appear at this order. 
	
\begin{figure}[!t]
		\begin{center}
			\includegraphics[width=0.7\textwidth]{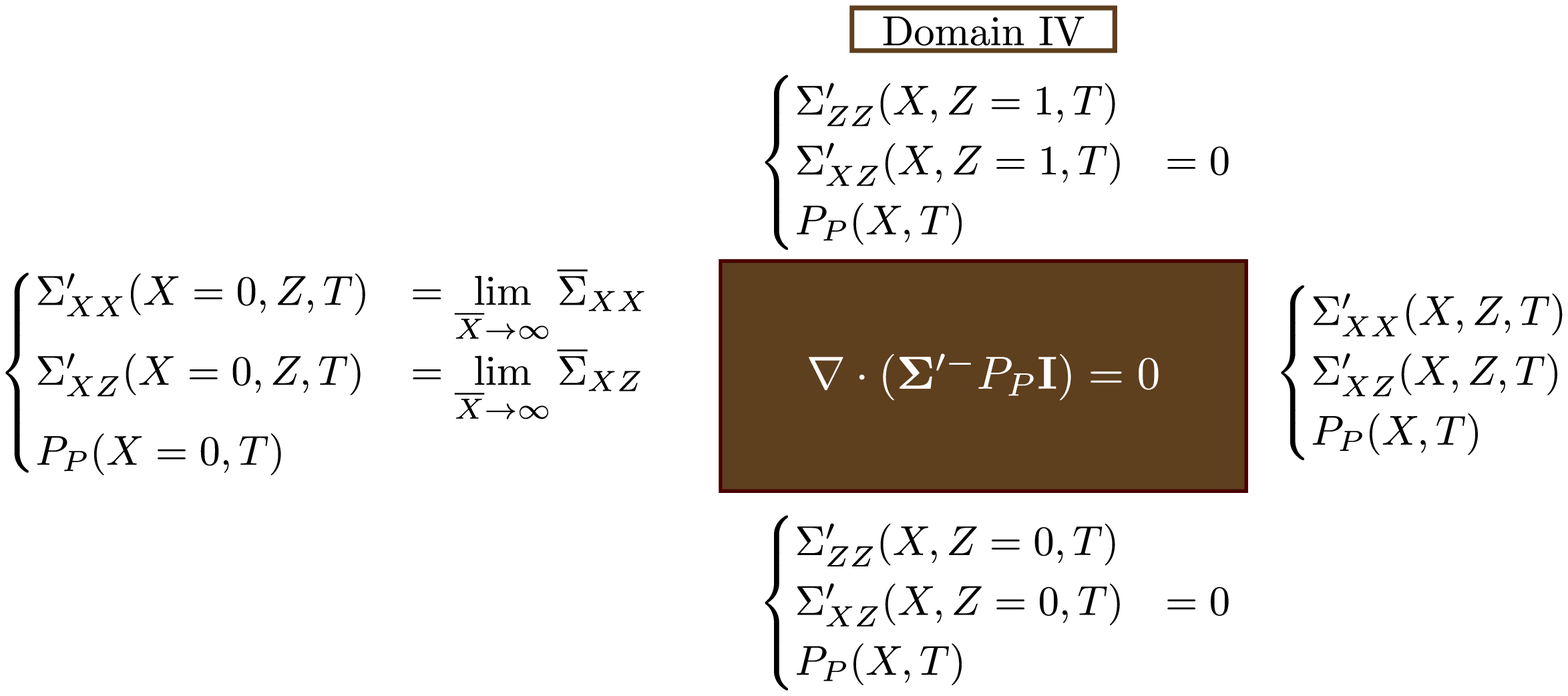}
		\end{center}
		
		\vspace{-4mm}
		\caption{Structure of the elastic problem in term of stresses in the outer poroelastic domain IV. On the left   the conditions are derived matching with the inner solution on domain III; on the top and bottom boundaries the conditions are obtained using the boundary conditions. On the right boundary we have the value for a generic coordinate $X$. }\label{fig:DivTheoremPorous}
	\end{figure}

	Before moving into the analysis of regions II and III, we determine the differential equations that must be satisfied by the functions $A_D^{(0)}(X,T)$ and $A_D^{(1)}(X,T)$.
	In domain IV we solve the equilibrium equation  $\nabla \cdot \left(\bSigma'-P_P \bI\right)=\bf{0}$ (see equation (\ref{EQ:balance_final})).  Denoting the area and boundary of the domain by $\mathcal{A}$ and $\mathcal{L}$, respectively, application of the divergence theorem to the square $\left[0,X\right] \times \left[0,1\right]$ shown in fig. \ref{fig:DivTheoremPorous} gives
	\begin{equation}\label{div}
	\int_{\mathcal{A}} \nabla \cdot \left(\bSigma'-P_P \bI\right) \upd \mathcal{A} = \int_{\mathcal{L}} \left(\bSigma'-P_P \bI\right)\cdot \nn \upd \mathcal{L} = \bf{0},
	\end{equation}
	where $\nn$ is the unit outward normal vector to the  boundary.  Equation (\ref{div}) then gives
	\begin{subequations}
		\begin{align}
		\label{EQ:div_x}
		& \int_0^1 \left(- \left(\Sigma'_{XX}-P_P\right)\big\vert_{X=0} + \left(\Sigma'_{XX}-P_P\right)\right) \upd \zeta = - \frac{1}{\epsilon}\int_0^X \left(\Sigma'_{XZ}\big\vert_{Z=1} -\Sigma'_{XZ}\big\vert_{Z=0} \right) \upd \xi=0,\\
		\label{EQ:div_z}
		&\int_0^1 \left(- \Sigma'_{XZ}\big\vert_{X=0} + \Sigma'_{XZ} \right) \upd \zeta = - \epsilon \int_0^X \left(\left(\Sigma'_{ZZ}-P_P\right)\big\vert_{Z=1} -\left(\Sigma'_{ZZ}-P_P\right)\big\vert_{Z=0} \right) \upd \xi.
		\end{align}
	\end{subequations}
	Equation \eqref{EQ:div_x} at leading-order gives 
	\begin{equation}
	\Sigma_{XX}'^{(0)} - P_P^{(0)} = \Sigma_{XX}'^{(0)}(X = 0, T) - P_P^{(0)}(X=0,T).
	\end{equation}
	Using the relationship
	(\ref{EQ:stress_outer}a), then gives an equation for  $A_D^{(0)}(X,T)$
	\begin{equation}
	\label{EQ:Evolutionary_condition_O1}
	\left(2 \cH_0^{(0)} + \cN_0^{(0)}\right) \frac{\partial A_D^{(0)}}{\partial X} - P_P^{(0)} = \Sigma_{XX}'^{(0)}\big \vert_{X = 0} - P_P^{(0)}(X=0,T).
	\end{equation}
	Equation \eqref{EQ:Evolutionary_condition_O1} is solved subjected to the  boundary condition $A_D^{(0)}(X=1,T)=0$ and the function $\Sigma_{XX}'^{(0)}\big \vert_{X = 0}(T)$ is unknown at this stage.
	
	At $O(\epsilon)$ analogous arguments can be used to obtain an equation for $A_D^{(1)}$ as
	\begin{equation}
	\label{EQ:Evolutionary_condition_Oeps}
	\left(2 \cH_0^{(0)} +\cN_0^{(0)}\right) \frac{\partial A_D^{(1)}}{\partial X} + \frac{\partial A_D^{(0)}}{\partial X}\int_0^1\left(2 \cH_0^{(1)} +\cN_0^{(1)}\right)\upd \zeta  - P_P^{(1)} = \Sigma_{XX}'^{(1)}\big \vert_{X = 0} - P_P^{{(1)}}(X=0,T).
	\end{equation}
	Equation \eqref{EQ:Evolutionary_condition_Oeps} is solved subjected to the boundary condition $A_D^{(1)}(X=1,T)=0$. Again, the function $\Sigma_{XX}'^{(1)}\big \vert_{X = 0}(T)$ is unknown.
	
	The leading-order solutions presented in \eqref{EQ:PorousD_Outer0}, \eqref{EQ:PorousV_Outer0} and \eqref{EQ:PorousV_Outer0}, equipped with the equation in \eqref{EQ:Evolutionary_condition_O1} for the function $A_D^{(0)}$, and the first order solutions presented in \eqref{EQ:PorousD_Outer1}, \eqref{EQ:PorousV_Outer1} and \eqref{EQ:PorousV_Outer1}, equipped with the equation in \eqref{EQ:Evolutionary_condition_Oeps} for the function $A_D^{(1)}$, are now determined up to four functions depending only on time $T$: $P_P^{01}(T)$, $P_P^{11}(T)$, $\Sigma_{XX}'^{(0)}\big \vert_{X = 0}(T)$ and $\Sigma_{XX}'^{(1)}\big \vert_{X = 0}(T)$. These four unknown functions are determined in  \S \ref{SECSEC:Transmission} when the coupling conditions across the outer domains are presented. In order to complete this step, however, we first need to study the structure of the inner domains II and III.
	
	\subsubsection{Fluid domain II and Poroelastic domain III}\label{sec:DomainsIIIII}
	
	At leading-order, the problems in the inner fluid domain \eqref{EQ:inner_fluid} and in the inner poroelastic domain \eqref{EQ:inner_porous} coupled via \eqref{EQ:inter-lubrication_inner} can be rearranged as follow. The leading-order pressure fields satisfy 
	\begin{align}\begin{cases}\label{EQ:Prex_inner0}
	\bar{\nabla} \oP_F^{(0)} = \cero, \quad  & \text{in region II},\\
	\bar{\nabla} \oP_P^{(0)} = \cero, \quad  & \text{in region III},\\
	\oP_P^{(0)} = \bar{K}\oP_F^{(0)}  \quad & \text{on $\oX = 0$},
	\end{cases}\end{align}
	which gives that  the pressures are constant and related via $\oP_P^{(0)}=\bar{K}\oP_F^{(0)}$ .
	
	The leading-order stress field in the poroelastic domain obeys
	\begin{equation}\label{EQ:Disp_inner0}\begin{cases}
	\bar{\nabla} \cdot  \obSigma'^{(0)} = \cero \quad  & \text{in region III},\\
	\bar{\Sigma}'^{(0)}_{XX} = \bar{\Sigma}'^{(0)}_{XZ} = 0  \quad &\text{on $\oX = 0$},\\
	\bar{\Sigma}'^{(0)}_{XZ} = \oD_Z^{(0)} = 0  \quad &\text{on $\oZ = 0,1$}.
	\end{cases}\end{equation}
	We first note that $\bar{\Sigma}'^{(0)}_{XX}$ is symmetric with respect to $\oZ = 0.5$. Applying the divergence theorem to equation (\ref{EQ:Disp_inner0})a in   $\left[0,\oX\right]\times \left[0,1\right]$, we find that $\int_0^1\bar{\Sigma}'^{(0)}_{XX}(\oX,Z,T)\upd \zeta=0$: thus $\bar{\Sigma}'^{(0)}_{XX} = 0$. If $\bar{\Sigma}'^{(0)}_{XX} = 0$, the axial component of (\ref{EQ:Disp_inner0}a), together with the boundary conditions $\bar{\Sigma}'^{(0)}_{XZ}=0$ at $\oZ=0,1$ gives   $\bar{\Sigma}'^{(0)}_{XZ}\equiv 0$. The transverse component of (\ref{EQ:Disp_inner0})a  then gives  $\bar{\Sigma}'^{(0)}_{ZZ}=h(\oX, T)$. Having determined the form of the stress components,  we use \eqref{EQ:StressesInner1} and \eqref{EQ:StressesInner2} to solve for the displacements: using 
	$\oD_Z^{(0)}(X,Z=0,1,T)=0$, we get $\oD^{(0)}_Z = 0$ (and $h(\oX, T)=0$) and $\oD^{(0)}_X = g(\oZ,T)$, \textit{i.e.}~$\oD^{(0)}_X$ is independent by $\oX$. Using \eqref{EQ:StressesInner3}, $g(\oZ,T)=g(T)$. The perhaps counterintuitive result is that the poroelastic inner region is stress-free and undergoes a rigid motion with $\obD^{(0)} = \left(\oD_X^{(0)}(T),0\right)$.

	At $O(\epsilon)$ equations \eqref{EQ:inner_fluid} and \eqref{EQ:inner_porous}, coupled via \eqref{EQ:inter-lubrication_inner}, reveal that the elastic problem decouples from the fluid flow at this order, and we can solve (i) an elastic problem for $\obD^{(1)}$ valid in region III only and (ii) a Stokes-Darcy problem in the unknowns $\left(\oP_F^{(1)}, \obU^{(0)},\oP_P^{(1)}\right)$ in regions II and III. 
	
	The elastic problem is
	\begin{align}\label{EQ:Disp_inner1}\begin{cases}
	\displaystyle \bar{\nabla} \cdot  \obSigma'^{(1)} = \cero \quad  & \text{in region III},\\
	\displaystyle \bar{\Sigma}'^{(1)}_{XX} = \oP_P^{(0)} - \bar{K}\oP_F^{(0)}, \quad \bar{\Sigma}_{XZ}^{'(1)} = 0 \quad & \text{on $\oX=0$},\\
	\bar{\Sigma}'^{(1)}_{XZ} = \oD_Z^{(1)}= 0  \quad &\text{on $\oZ = 0,1$}.
	\end{cases}\end{align} 
	Recalling that $\oP_P^{(0)} - \bar{K}\oP_F^{(0)} = 0$ at $\oX=0$ (see equation \ref{EQ:Prex_inner0}c), we can repeat the arguments made at leading-order  to show that, at this order, the poroelastic domain is again stress-free and undergoes a displacement $\obD^{(1)} = \left(\oD_X^{(1)}(T),0 \right)$. 
	
	The coupled Stokes-Darcy problem is
	\begin{align}\label{EQ:StokesDarcy_Inner}\begin{cases}
	\nabla^2 \obU^{(0)} - \nabla \oP_F^{(1)} = \cero  \quad \nabla \cdot \obU^{(0)}  = 0, \quad  & \text{in region II},\\
	\obV^{(0)} = -\cK_0^{(0)}(0) \nabla  \oP_P^{(1)} \quad  \nabla \cdot \obV^{(0)} = 0, \quad  & \text{in region III},\\
	\displaystyle \oU_{X}^{(0)} =\frac{\partial \oD_X^{(0)}}{\partial T} -\cK_0^{(0)}(0) \frac{\partial \oP_P^{(1)}}{\partial X} \quad &\text{on $\oX=0$},\\
	\displaystyle  \oP_P^{(1)}  = \bar{K}\oP_F^{(1)} - 2 \bar{K} \frac{\partial \oU_{X}^{(0)}}{\partial X}\quad &\text{on $\oX=0$},\\ \displaystyle{\frac{\partial \oU_{X}^{(0)}}{\partial Z} + \frac{\partial \oU_{Z}^{(0)}}{\partial X}  = \gamma \frac{1}{\sqrt{\bar{K}\cK_0^{(0)}(0)}}\oU_{Z}^{(0)}} \quad &\text{on $\oX=0$}.
	\end{cases}\end{align}
	This system is solved subject to the following far field conditions, acting as boundary conditions at the inlet of domain II and the outlet of domain III, obtained from matching the inner solutions to the solutions in the outer domain, via
	\begin{align}
	&\obU^{(0)} \sim 6 \oZ \left(1-\oZ\right)f(T) \be_X \quad & \text{for $\oX \rightarrow - \infty$},\\
	&\oP_P^{(1)} \sim \left(P_P^{(1)}(0,T) + \frac{\partial P_P^{(0)}(\epsilon \oX,T)}{\partial \epsilon} \big\vert_{\epsilon=0}\right) \quad & \text{for $\oX \rightarrow + \infty$}.
	\end{align}
	This problem can be solved numerically to obtain the leading-order velocities fields $\obU^{(0)}$, $\obV^{(0)}$ and the first order pressure fields $\oP_F^{(1)}$, $\oP_P^{(1)}$.

	\subsubsection{Coupling conditions across the outer domains} \label{SECSEC:Transmission}
	The analysis in the outer domains, considering both the fluid and the poroelastic ones, left us with six unknowns to be determined (three for the leading-order problem and three for order $O(\epsilon)$ problem): $P_F^{00}(T)$ and $P_F^{10}(T)$ for the outer fluid domain and  $P_P^{01}(T)$, $P_P^{11}(T)$, $\Sigma_{XX}'^{(0)}\big \vert_{X=0}(T)$ and $\Sigma_{XX}'^{(1)}\big \vert_{X=0}(T)$ for the outer poroelastic domain. To close the problem, we thus need six coupling  conditions, imposed at $X=0$, to relate the outer solutions in domains I and  IV.  These effective coupling conditions result from  matching the solutions in the inner domains II and III with the outer solutions in domains I and IV.
	
	From \eqref{EQ:Prex_inner0} we found that the pressure is constant across the inner domains, with $\bar{K}\oP_F^{(0)} = \oP_P^{(0)}$. Matching the inner solutions with the outer domains, we get $\oP_P^{(0)} \sim P_P^{(0)}(0,T)$ for $\oX \rightarrow + \infty$ and $\oP_F^{(0)} \sim P_F^{(0)}(0,T)$ for $\oX \rightarrow - \infty$. From \eqref{EQ:Fluid_Outer0}, $P_F^{(0)}(T) =  P_F^{00}(T)$ so the coupling condition for the leading-order pressures is
	\begin{equation}
	\label{EQ:int_P0}
	P_F^{00} = \frac{P_P^{(0)}(0,T)}{\bar{K}}.
	\end{equation}
	At the order $O(\epsilon)$ the inner domains are governed by the Stokes-Darcy problem in \eqref{EQ:StokesDarcy_Inner}; once $\oP_F^{(1)}$ is computed throughout domain II, we can use the matching condition with the outer fluid domain $\oP_F^{(1)} \sim P_F^{10} - 12 \oX f(T)$ for $\oX \rightarrow -\infty$ to compute $P_F^{10}(T)$. 
	
	From problems \eqref{EQ:Disp_inner0} and \eqref{EQ:Disp_inner1} we see that the displacement in the inner domain is in the axial direction and  is spatially homogeneous at both the leading and first order. Domain III is also stress free at these orders, and we can match the inner stress field with outer stress field for $\oX \rightarrow \infty$ to obtain
	\begin{equation}\label{EQ:int_stress}\begin{split}
	\lim\limits_{\oX \rightarrow \infty} \bar{\Sigma}'^{(0)}_{XX} = 0 =  \Sigma'^{(0)}_{XX}\big\vert_{X=0}, \quad \lim\limits_{\oX \rightarrow \infty} \bar{\Sigma}'^{(0)}_{XZ} = 0 =  \Sigma'^{(0)}_{XZ}\big\vert_{X=0},\\
	\lim\limits_{\oX \rightarrow \infty} \bar{\Sigma}'^{(1)}_{XX} = 0 =  \Sigma'^{(1)}_{XX}\big\vert_{X=0}, \quad \lim\limits_{\oX \rightarrow \infty} \bar{\Sigma}'^{(1)}_{XZ} = 0 =  \Sigma'^{(1)}_{XZ}\big\vert_{X=0},
	\end{split}
	\end{equation} 
	from which we obtain the unknowns $\Sigma'^{(0)}_{XX}\big\vert_{X=0}$ and $\Sigma'^{(1)}_{XX}\big\vert_{X=0}$ required for  \eqref{EQ:Evolutionary_condition_O1} and \eqref{EQ:Evolutionary_condition_Oeps}.
	
\begin{figure}[!t]
		\begin{center}
			\includegraphics[width=0.8\textwidth]{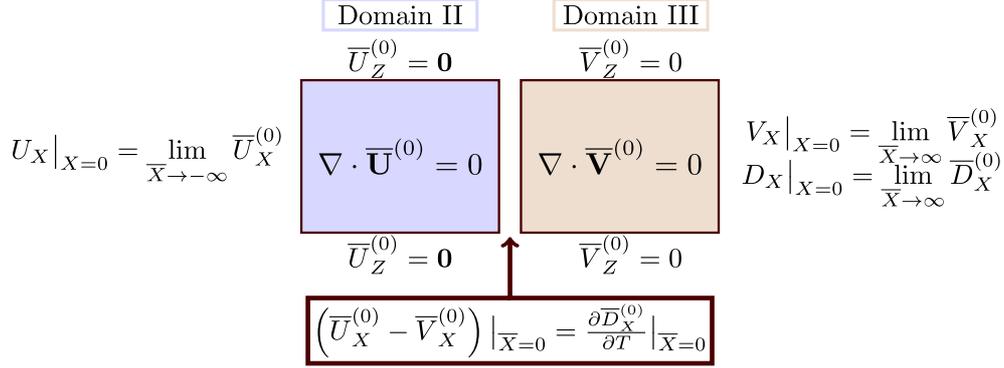}
		\end{center}
		
		\vspace{-4mm}
		\caption{Leading-order structure of  velocity in the inner domains II and III. Matching conditions with the outer solutions are shown on the left of domain II and on the right of domain III.}\label{fig:DivTheoremVelocities}
	\end{figure}

	To obtain $P_P^{01}(T)$ and $P_P^{11}(T)$, we use a global conservation of mass argument \cite{Dalwadi2016}. First, we focus on the leading-order and we refer to the problem for the velocities as shown in fig. \ref{fig:DivTheoremVelocities}. In the inner domains, from \eqref{EQ:StokesDarcy_Inner}, we know that $\nabla \cdot \obU^{(0)} =0$ in region II and $\nabla \cdot \obV^{(0)}=0$ in region III; applying the divergence theorem to the inner domains and using the boundary conditions for the normal velocities at $Z=0,1$ in \eqref{EQ:BCS_inner}, we can write
	\begin{equation}
	\int_0^1 \left(\oU_{X}^{(0)} - \oV_{X}^{(0)}\right)\big\vert_{\oX=0} \upd \zeta = \int_0^1 \lim\limits_{\oX \rightarrow -\infty}\oU_{X}^{(0)} \upd \zeta -\int_0^1 \lim\limits_{\oX \rightarrow \infty} \oV_{X}^{(0)} \upd \zeta.
	\end{equation}
	Employing 
	\eqref{eq:inter-U_lubrication_inner} across domains II and III at leading-order $\left(\oU_{X}^{(0)} - \oV_{X}^{(0)}\right)\big\vert_{\oX=0} = \partial \oD^{0}_X/ \partial T \big\vert_{\oX =0}$, and the matching conditions for the velocity field in the fluid domain, $\oU_{X}^{(0)} \big\vert_{\oX \rightarrow -\infty} =  U_{X}^{(0)} \big\vert_{X =0}$, and for the velocity and displacement fields in the poroelastic domain, $ \lim\limits_{\oX \rightarrow \infty} \oV_{X}^{(0)} =  V_{X}^{(0)} \big\vert_{X =0}$ and $\oD_X^{(0)}\big\vert_{\oX=0} = \lim\limits_{\oX \rightarrow \infty}\oD_X^{(0)} = D_X^{(0)}\big\vert_{X=0}$, gives the following coupling condition for the outer I and IV domains
	\begin{equation}\label{EQ:int_V0}
	\int_0^1 U_{X}^{(0)}(X=0,T) \upd \zeta = \int_0^1\left(\frac{\partial D_X^{(0)}}{\partial T}\big \vert_{X=0} + V_X^{(0)}(X=0,T)\right)\upd \zeta,
	\end{equation}
	where we point out that the function $P_P^{01}(T)$ is in the expression of $V_X^{(0)}(X,T)$. Analogous arguments can be used at order $O(\epsilon)$ to obtain the function $P_P^{11}(T)$ using
	\begin{align}\label{EQ:int_V1}
	\int_0^1 U_{X}^{(1)}(X=0,T) \upd \zeta &= \int_0^1\left(\frac{\partial D_X^{(1)}}{\partial T}\big \vert_{X=0} + V_X^{(1)}(X=0,T)\right)\upd \zeta.
	\end{align}
	
	\subsubsection{Solutions in the poroelastic domain IV}\label{sec:solutionIV}
	Employing the coupling conditions  \eqref{EQ:int_stress}, \eqref{EQ:int_V0} and \eqref{EQ:int_V1}, we can state the full solution for the pressure, displacement and velocities in  domain IV. At leading-order,  we obtain
	\begin{subequations}\label{eq:quartet0}
		\begin{align}
		\label{EQ:PP0}
		&P_P^{(0)}(X,T) = -f(T)\int_1^X \frac{1}{\cK_0^{(0)}(\xi)}\upd \xi + \int_1^X  \frac{\partial  A_D^{(0)}(\xi,T)}{\partial T} \frac{1}{\cK_0^{(0)}(\xi)}\upd \xi,\\
		\label{EQ:Vd0}
		&\bV^{(0)}(X,T) = \left(f(T) - \frac{\partial A_D^{(0)}(X,T)}{\partial T} \right) \mathbf{e}_X,\\
		\label{EQ:D0}
		&\bD^{(0)}(X,T) = A_D^{(0)}(X,T) \mathbf{e}_X,
		\end{align}\end{subequations}
	with $A_D^{(0)}(X,T)$ given by the solution of \eqref{EQ:Evolutionary_condition_O1}. At leading-order, the pressure, displacement and velocity fields do not depend on the transverse coordinate $Z$. 
	
	At first order, the solution is
	\begin{subequations}\label{eq:quartet1}
		\begin{align}
		\label{EQ:PP1}
		P_P^{(1)}(X,T) & = \int_1^X \frac{\partial A_D^{(1)}(\xi,T)}{\partial T}\frac{1}{\cK_0^{(0)}(\xi)}\upd \xi \nonumber \\
		&\qquad +\int_1^X \frac{1}{\cK_0^{(0)}(\xi)}\left(-f(T) +\frac{\partial A_D^{(0)}(\xi,T)}{\partial T}\right)\left( \int_0^1 \frac{\cK_0^{(1)}(\xi,\zeta)}{\cK_0^{(0)}(\xi)}\upd \zeta\right) \upd \xi,\\
		\label{EQ:Vd1}
		\bV^{(1)}(X,Z,T) & =\left(-\frac{\partial A_D^{(1)}}{\partial T} + \left(f(T) - \frac{\partial A_D^{(0)}}{\partial T}\right) \left(\frac{\cK_0^{(1)}(X,Z)}{\cK_0^{(0)}(X)} - \int_0^1\frac{\cK_0^{(1)}(X,Z)}{\cK_0^{(0)}(X)} \upd Z\right) \right)\mathbf{e}_X \nonumber\\
		&\qquad + \left[\int_0^Z \frac{\partial}{\partial X}\left(\left(-f(T) + \frac{\partial A_D^{(0)}(X,T)}{\partial T} \right)\frac{\cK_0^{(1)}(X,\zeta)}{\cK_0^{(0)}(X)}\right)\upd \zeta \right. \nonumber\\
		&\qquad \qquad \left. - Z \frac{\partial}{\partial X}\left(\left(-f(T) + \frac{\partial A_D^{(0)}(X,T)}{\partial T} \right)\int_0^Z\frac{\cK_0^{(1)}(X,\zeta)}{\cK_0^{(0)}(X)}\upd \zeta\right)\right] \mathbf{e}_Z,\\
		\label{EQ:D1}
		\bD^{(1)}(X,Z,T) & = A_D^{(1)}(X,T) \mathbf{e}_X \nonumber\\
		&\quad + \frac{1}{2 \cH_0^{(0)}(X)+\cN_0^{(0)}(X)}\left(\! Z\!\!\int_0^1\!\!\cN_0^{(1)}(X,\zeta)\upd \zeta-\int_0^Z \!\!\cN_0^{(1)}(X,\zeta)\upd \zeta\right)\frac{\partial A_D^{(0)}}{\partial X} \mathbf{e}_Z,
		\end{align}\end{subequations}
	with $A_D^{(1)}(X,T)$ given by the solution of \eqref{EQ:Evolutionary_condition_Oeps}. We see that the first order correction to  the pressure $P_P^{(1)}$ does not depend on $Z$, and only retains the $X$-dependence. The Darcy velocity and the displacement field, however, do vary in $Z$: specifically, this dependence is controlled by the expression of $\cK_0^{(1)}$ for the former and by $\cH_0^{(1)}$ for the latter. 
	
	\subsection{Limiting cases}\label{sec:limitingcases}
	Having determined analytical expressions for the flow and displacement in the outer domain IV, we now consider some limiting cases. 
	
	\noindent {\bf Steady flow.} If the driving upstream flow is steady  ($f(T)=1$) all the time derivatives appearing in \eqref{eq:quartet0} and \eqref{eq:quartet1} are zero. The solutions at leading-order are then
	\begin{subequations}\begin{align}
		\label{EQ:PP0_stat}
		&P_P^{(0)}(X) = -\int_1^X \frac{1}{\cK_0^{(0)}(\xi)}\upd \xi,\\
		\label{EQ:Vd0_stat}
		&\bV^{(0)}(X) = \mathbf{e}_X,\\
		\label{EQ:D0_stat}
		&\bD^{(0)}(X) = \int_{X}^1\left[\frac{1}{\left(2\cH_0^{(0)}(\xi)+\cN_0^{(0)}(\xi)\right)}\int_0^{\xi}\frac{1}{\cK_0^{(0)}(\eta)}\upd \eta\right] \upd \xi \mathbf{e}_X.
		\end{align}\end{subequations}
	while at order $O(\epsilon)$  we have
	{\small\begin{subequations}
			\begin{align}
			\label{EQ:PP1_stat}
			P_P^{(1)}(X) & = -\int_1^X \frac{1}{\cK_0^{(0)}(\xi)}\left( \int_0^1 \frac{\cK_0^{(1)}(\xi,\zeta)}{\cK_0^{(0)}(\xi)}\upd \zeta\right) \upd \xi,\\
			\label{EQ:Vd1_stat}	
			\bV^{(1)}(X,Z) & =\left( \left(\frac{\cK_0^{(1)}(X,Z)}{\cK_0^{(0)}(X)} - \int_0^1\frac{\cK_0^{(1)}(X,Z)}{\cK_0^{(0)}(X)} \upd Z\right) \right)\mathbf{e}_X \nonumber\\
			&\qquad + \left[-\int_0^Z \frac{\partial}{\partial X}\left(\frac{\cK_0^{(1)}(X,\zeta)}{\cK_0^{(0)}(X)}\right)\upd \zeta + Z \frac{\partial}{\partial X}\left(\int_0^Z\frac{\cK_0^{(1)}(X,\zeta)}{\cK_0^{(0)}(X)}\upd \zeta\right)\right] \mathbf{e}_Z,\\ 
			\bD^{(1)}(X,Z) & = \biggl[-\int_X^1 \left(\frac{1}{2\cH_0^{(0)}(\xi) + \cN_0^{(0)}(\xi)}\right)\left( \int_0^{\xi}\frac{1}{\cK_0^{(0)}(\xi)}\left( \int_0^1 \frac{\cK_0^{(1)}(\xi,\zeta)}{\cK_0^{(0)}(\xi)}\upd \zeta\right) \upd \eta\right)\upd \xi \nonumber\\
			& \quad +\int_X^1\left(\frac{\int_0^1\left( 2\cH_0^{(1)}(\xi,\zeta)+\cN_0^{(1)}(\xi,\zeta)\right)\upd \zeta}{\left(2 \cH_0^{(0)}(\xi)+\cN_0^{(0)}(\xi)\right)^2}\int_0^{\xi} \frac{1}{\cK_0^{(0)}(\xi)}\upd \eta \right)\upd \xi \biggr] \mathbf{e}_X \nonumber\\
			&\quad -\!(2 \cH_0^{(0)}(X)+\cN_0^{(0)}(X))^{-2}\!\!\int_0^X \frac{1}{\cK_0^{(0)}(\xi)}\upd \xi\left(\! Z\!\!\int_0^1\!\!\cN_0^{(1)}(X,\zeta)\upd \zeta-\int_0^Z \!\!\cN_0^{(1)}(X,\zeta)\upd\! \zeta\right) \mathbf{e}_Z.\label{EQ:D1_stat}
			\end{align}
	\end{subequations}}
	In this regime,  the deformation of the solid skeleton and the flow of the fluid phase decouple. Indeed,  pressure and   velocity   are uniquely determined by the permeability function $\cK_0$, while the elastic properties $\cH_0$ and $\cN_0$ only enter in the expression for  displacement. 
	
	\noindent{\bf Rigid solid skeleton.} Assuming rigid solid skeleton implies $A_D^{(0)}, A_D^{(1)} \rightarrow 0$ and the the sets of solutions \eqref{EQ:PP0}, \eqref{EQ:Vd0} and \eqref{EQ:PP1}, \eqref{EQ:Vd1} can be simplified as
	\begin{subequations}
		\begin{align}
		\label{EQ:PP0_rigid}
		&P_P^{(0)}(X,T) = -f(T)\int_1^X \frac{1}{\cK_0^{(0)}(\xi)}\upd \xi,\\
		\label{EQ:Vd0_rigid}
		&\bV^{(0)}(X,T) = f(T) \mathbf{e}_X,
		\end{align}\end{subequations}
	and
	\begin{subequations}
		\begin{align}
		\label{EQ:PP1_rigid}
		P_P^{(1)}(X,T) & = - f(T)\int_1^X \frac{1}{\cK_0^{(0)}(\xi)}\left( \int_0^1 \frac{\cK_0^{(1)}(\xi,\zeta)}{\cK_0^{(0)}(\xi)}\upd \zeta\right) \upd \xi,\\
		\bV^{(1)}(X,Z,T) & =f(T) \left(\frac{\cK_0^{(1)}(X,Z)}{\cK_0^{(0)}(X)} - \int_0^1\frac{\cK_0^{(1)}(X,Z)}{\cK_0^{(0)}(X)} \upd Z\right)\mathbf{e}_X \nonumber\\
		\label{EQ:Vd1_rigid}
		\qquad &+ f(T) \left[- \int_0^Z \frac{\partial}{\partial X}\left(\frac{\cK_0^{(1)}(X,\zeta)}{\cK_0^{(0)}(X)}\right)\upd \zeta  + Z \frac{\partial}{\partial X}\left(\int_0^Z\frac{\cK_0^{(1)}(X,\zeta)}{\cK_0^{(0)}(X)}\upd \zeta\right)\right] \mathbf{e}_Z, 
		\end{align}\end{subequations}
	showing that any temporal dynamics for the pressure and the Darcy velocity, at both orders in the rigid porous domain, is proportional to the imposed  $f(T)$ at the inlet.
	
	\noindent{\bf Homogeneous material properties.}
	In the limit of homogeneous material properties we can write $\cK_0 (X,Z) = \bar{\cK}_0$,  $\cH_0 (X,Z) = \bar{\cH}_0$ and  $\cN_0 (X,Z) = \bar{\cN}_0$. At  $O(1)$, the solutions  \eqref{eq:quartet0} simplify as
	\begin{subequations}
		\begin{align}
		\label{EQ:PP0_homo}
		&P_P^{(0)}(X,T) = \frac{1}{\bar{\cK}_0} \left(-f(T)\left(X-1\right) + \int_1^X\frac{\partial  A_D^{(0)}(\xi,T)}{\partial T}\upd \xi\right),\\
		\label{EQ:Vd0_homo}
		&\bV^{(0)}(X,T) = \left(f(T) - \frac{\partial A_D^{(0)}(X,T)}{\partial T} \right) \mathbf{e}_X,\\
		\label{EQ:D0_homo}
		&\bD^{(0)}(X,T) = A_D^{(0)}(X,T) \mathbf{e}_X,
		\end{align}\end{subequations}
	with $A_D^{(0)}(X,T)$ being the solution of the equation \eqref{EQ:Evolutionary_condition_O1} with constant elastic parameters. At the order $O(\epsilon)$, the solutions \eqref{EQ:PP1}, \eqref{EQ:Vd1} and \eqref{EQ:D1} rewrite as 
	\begin{subequations}
		\begin{align}
		\label{EQ:PP1_homo}
		P_P^{(1)}(X,T) & = \frac{1}{\bar{K}\bar{\cK}_0}\int_1^X \frac{\partial A_D^{(1)}(\xi,T)}{\partial T}\upd \xi,\\
		\label{EQ:Vd1_homo}
		\bV^{(1)}(X,Z,T) & =\cero,\\
		\label{EQ:D1_homo}
		\bD^{(1)}(X,T) & = A_D^{(1)}(X,T) \mathbf{e}_X, 
		\end{align}\end{subequations}
	with $A_D^{(1)}(X,T)$ now solution of
	\begin{equation}
	\label{EQ:Evolutionary_condition_oeps}
	-P_P^{1}(X=0,T) = \left(2 \bar{\cH}_0 +\bar{\cN}_0\right) \frac{\partial A_D^{(1)}}{\partial X}  - P_P^{(1)}.
	\end{equation}
	In the homogeneous case,  we lose any dependence of the solutions on the axial coordinate, and the correction to the Darcy velocity $\bV^{(1)}(X,Z,T) $ is zero.
	
	\section{Results} \label{sec:result}
	With the solutions presented in the previous section, we now move into validating and discussing our findings with the aim to highlight the effect that inhomogeneous properties and deformability have on the mechanics of the outer porous domain, when coupled to the fluid domain.
	\subsection{Finite element method}\label{sec:fem}
	The finite element method that we use here to validate our solutions is analysed in \cite{ruiz21} and it solves for the poroelasticity equations in $\Omega_P$, in terms of the solid displacement $\bd$, the fluid  pressure $p_P$, and the total pressure $\varphi=  p_P - \lambda\nabla \cdot \bd$,  as 
	\begin{subequations}
		\begin{align}\label{eq:momM}
		-\nabla \cdot (2\hat{\mu}_s\cH\be(\bd) -\varphi \bI)& = \cero, & \\
		\varphi - p_P + \hat{\mu}_s\cN \left(\nabla \cdot \bd\right) & = 0, & \\
		\label{eq:massM}
		\frac{1}{\lambda}\frac{\partial p_P}{\partial t} - \frac{1}{\lambda} \frac{\partial \varphi}{\partial t}  - \nabla \cdot \biggl(\frac{\hat{\kappa}\cK}{\mu_f} \nabla p_P\biggr) & = 0. & 
		\end{align}\end{subequations}
	The finite element scheme is of second-order. It uses Taylor-Hood elements: overall continuous and piecewise quadratic polynomials for velocity, displacement, and fluid pressure approximation; and overall continuous and piecewise affine polynomials for the Stokes and poroelastic total pressure. The discretisation in time follows a backward Euler method with constant time step $\Delta t$.
	
	\subsection{Analysis and validation}
\begin{figure}[!t]
		\begin{center}
			\includegraphics[width=0.325\textwidth]{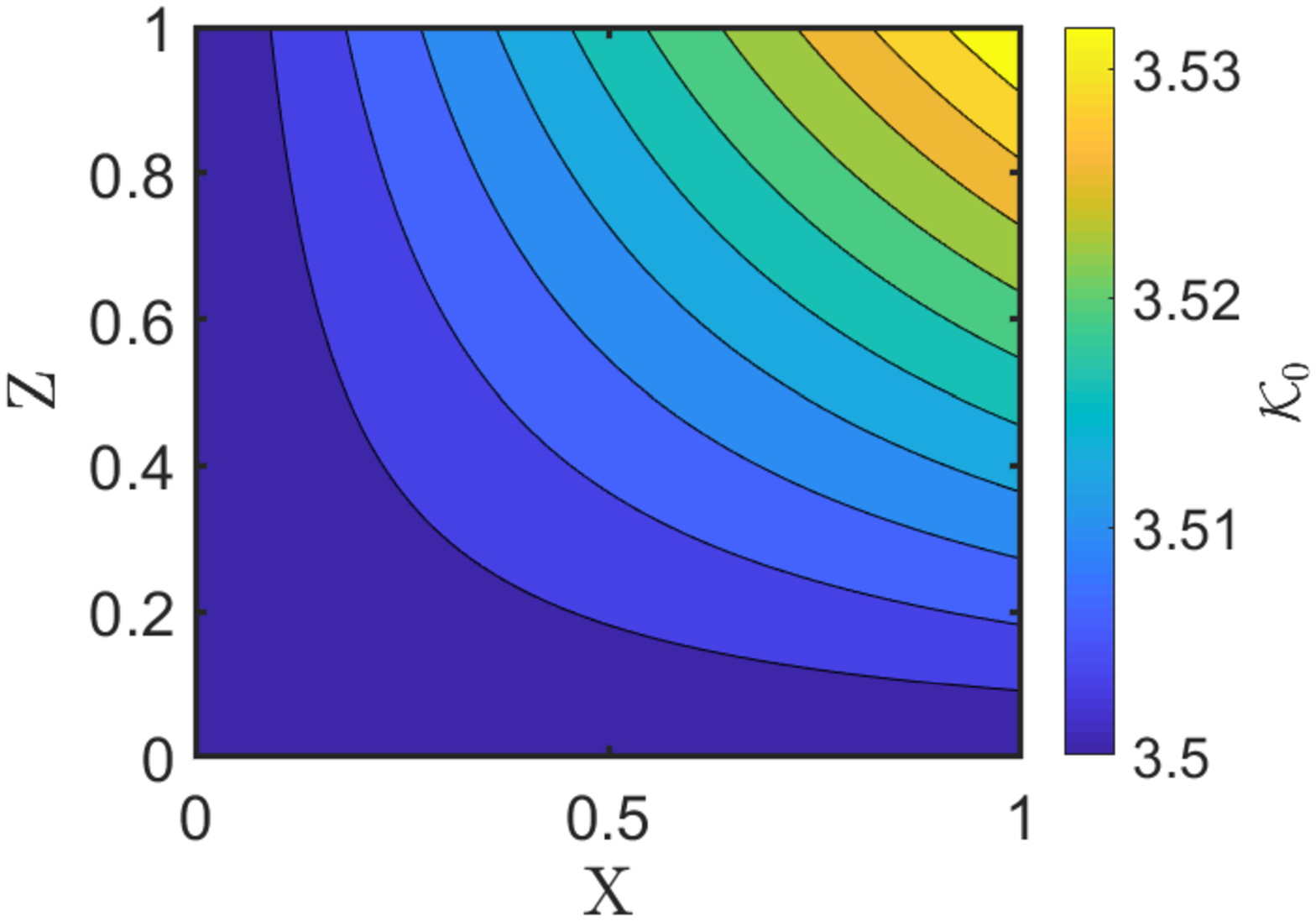}
			\includegraphics[width=0.325\textwidth]{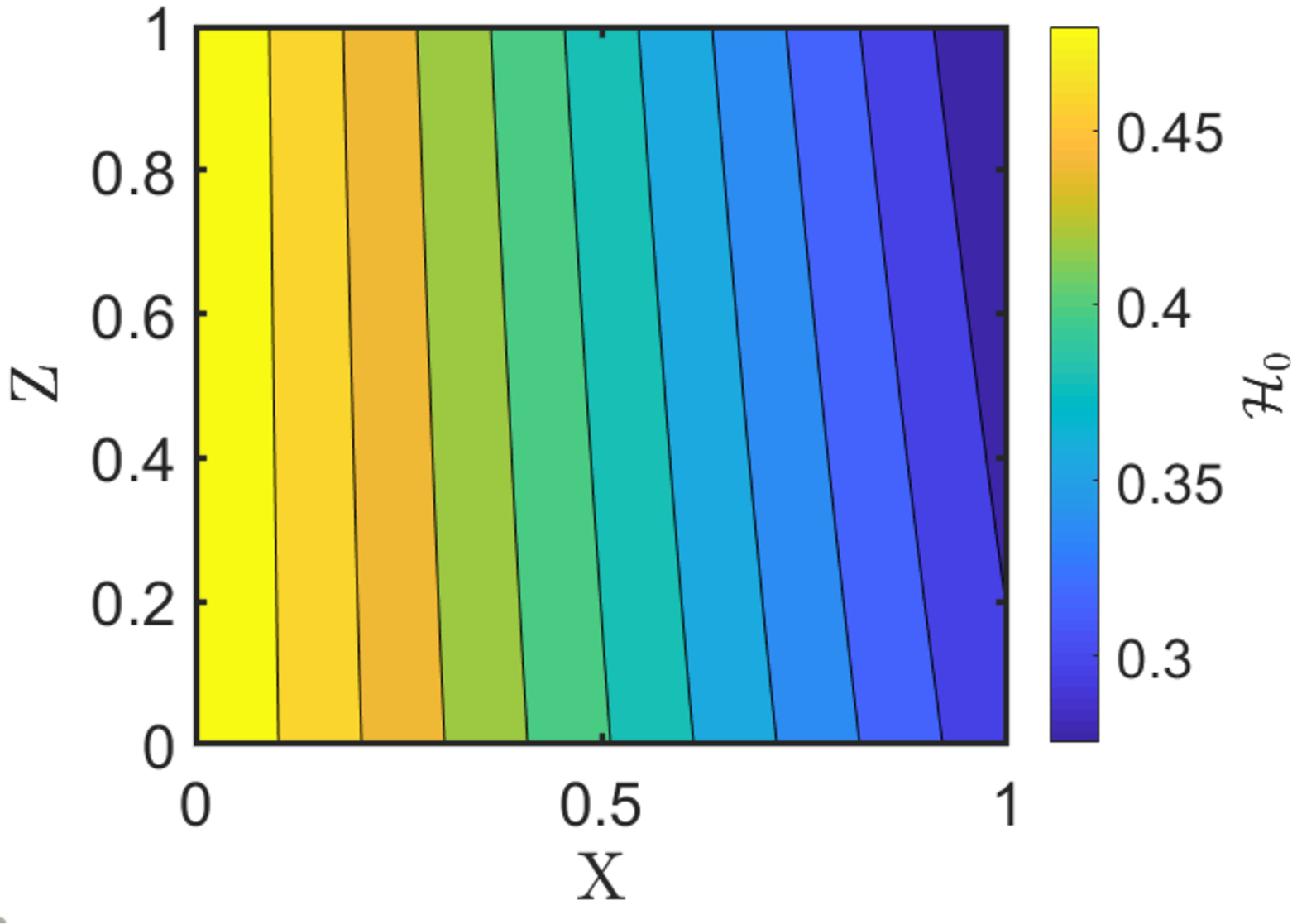}
			\includegraphics[width=0.325\textwidth]{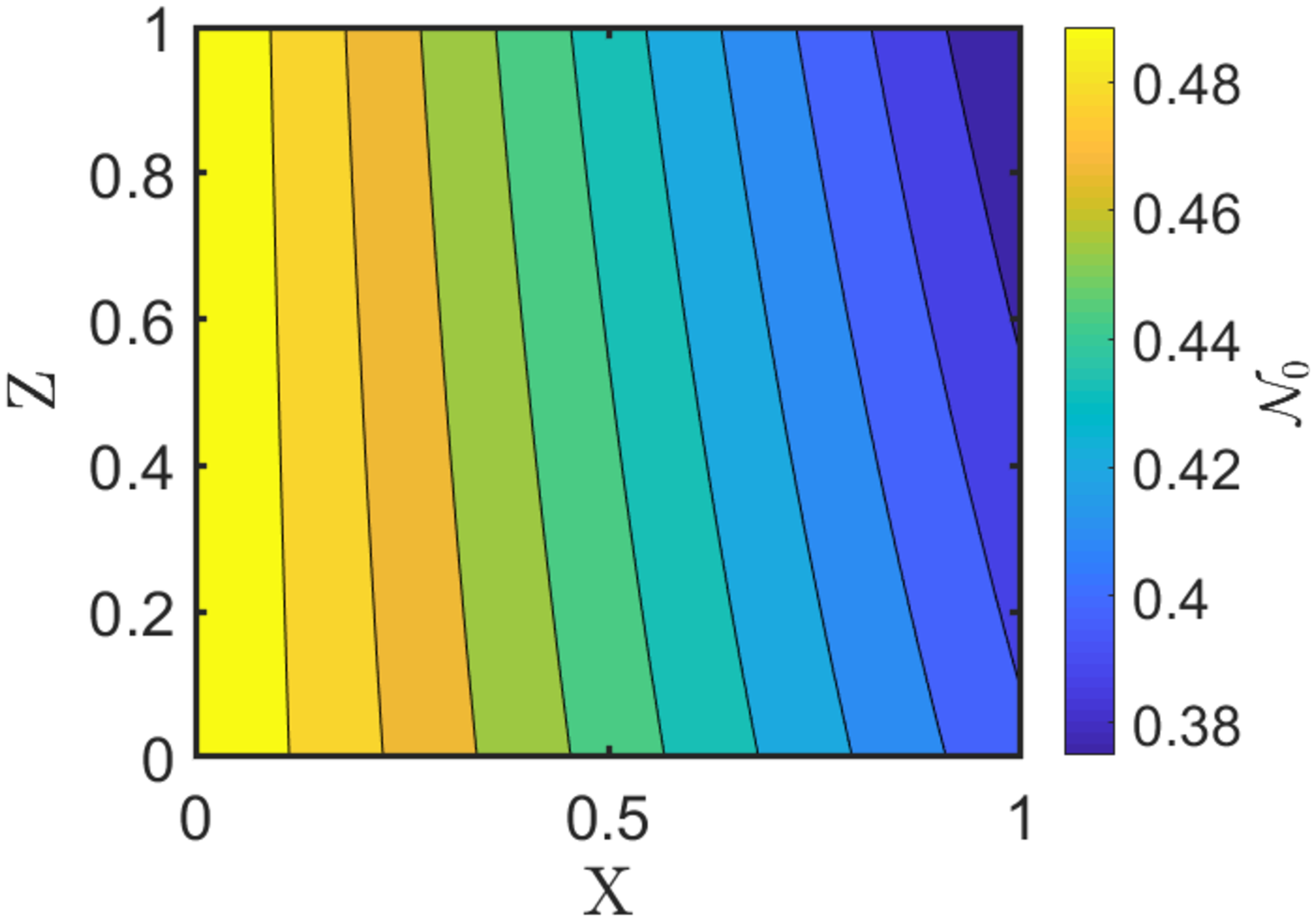}
		\end{center}
		
		\vspace{-4mm}
		\caption{Spatial distribution in the poroelastic domain of $\cK_0$ (left), $\cH_0$ (center) and $\cN_0$ (right).}\label{fig:maps}
	\end{figure}

	We motivate our choice of heterogeneous material properties in the poroelastic domain by the following observations. From \eqref{EQ:PP0}, \eqref{EQ:Vd0} and \eqref{EQ:D0} we see that, at leading-order, everything depends only on $X$. Analyzing the term in the parenthesis for the transverse component in \eqref{EQ:D1}, we note that a cross term of the type $\left(X Z\right)$ in $\cN^{(1)}$ allows us to have the transverse component of the displacement at $O(\epsilon)$ to be nonzero and to depend on both $X$ and $Z$, thus giving rise to a nonzero shear component of the stress tensor. From \eqref{EQ:Vd1}, considering the term in the square brackets, we see that a nonzero transverse Darcy velocity at $O(\epsilon)$ can be obtained if $\cK_0^{(1)}$ is either of the type $\left(X Z\right)$ or depending on $(Z)$ only. We therefore  prescribe the following functional forms for the material properties
	\begin{align}
	\label{eq:choice}
	\cK_0 = 1.5 + 2.0 X + \epsilon 7.0 X Z,\,  \cH_0 = 0.5 - 0.2 X - \epsilon 5 X Z,\,  \cN_0 = 0.5 -0.1 X - \epsilon 5 X Z,
	\end{align}
	and we plot their maps in the (dimensionless) poroelastic domain as shown in fig. \ref{fig:maps}. With this choice, the permeability increases moving from inlet to outlet and from bottom to top. The elastic properties show, instead, an opposite trend. This is consistent with what we would have expected if they were linked to a spatially-varying porosity field, instead of being prescribed directly on the spatial coordinates $X$ and $Z$ for simplicity: indeed, a larger porosity is related to a larger permeability and smaller effective elastic properties.
	Since we want to compare the heterogeneous elastic domain also against the homogeneous case, we prescribe the properties in this latter case by computing the mean value of the leading-order terms in \eqref{eq:choice} as
	\begin{align}
	\bar{\cK}_{0} = \int_{0}^1 \cK_0^{(0)} \upd X =2.5,\,  \bar{\cH}_{0} = \int_{0}^1 \cH_0^{(0)} \upd X =0.4,\,  \bar{\cN}_{0} = \int_{0}^1 \cN_0^{(0)} \upd X =0.45.
	\end{align}
	We finally set $\bar{K} = 0.06$ and the aspect ratio $\epsilon = 0.005$. 
	
	\begin{figure}[!t]
		\begin{center}
			\includegraphics[width=0.325\textwidth]{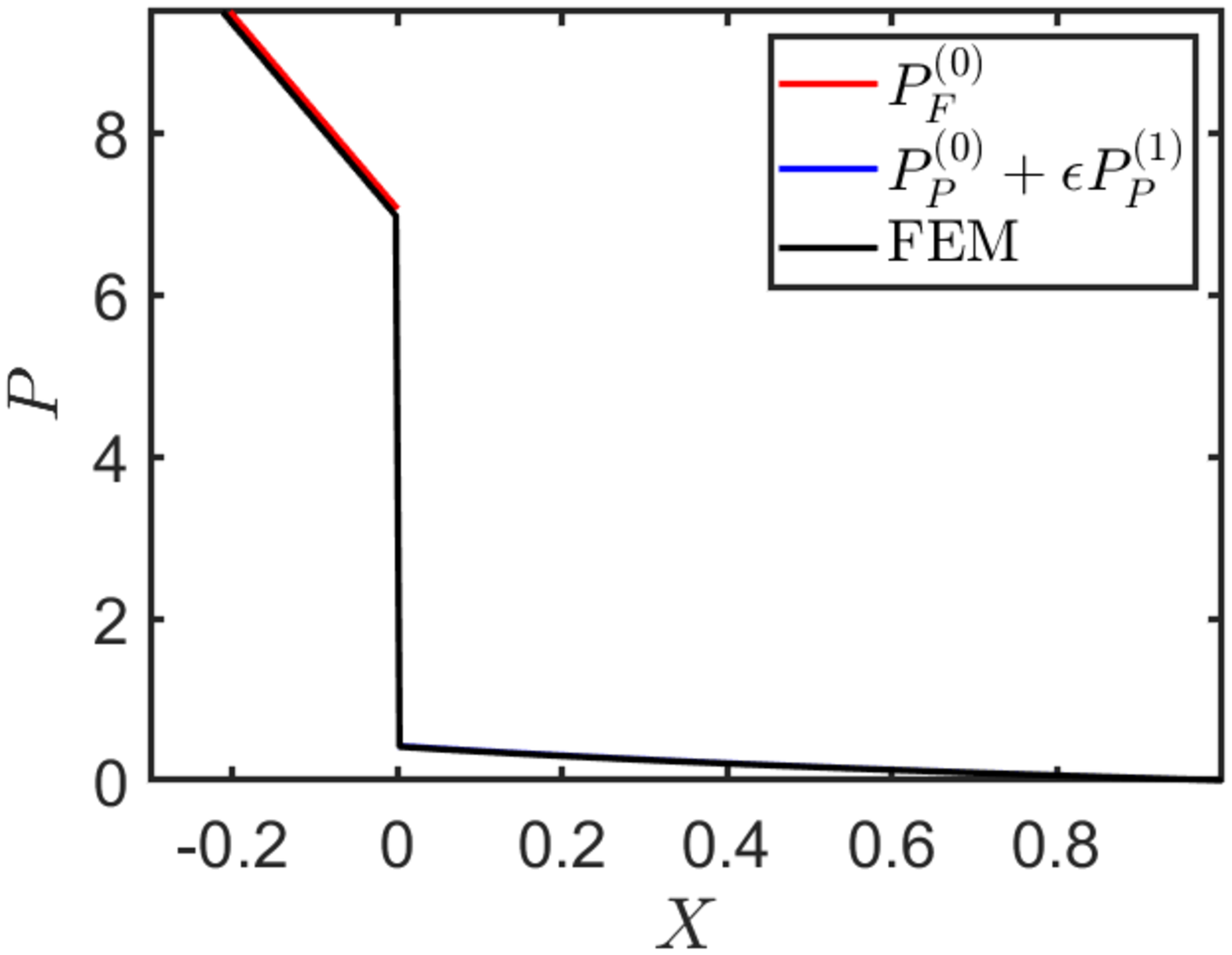}
			\includegraphics[width=0.325\textwidth]{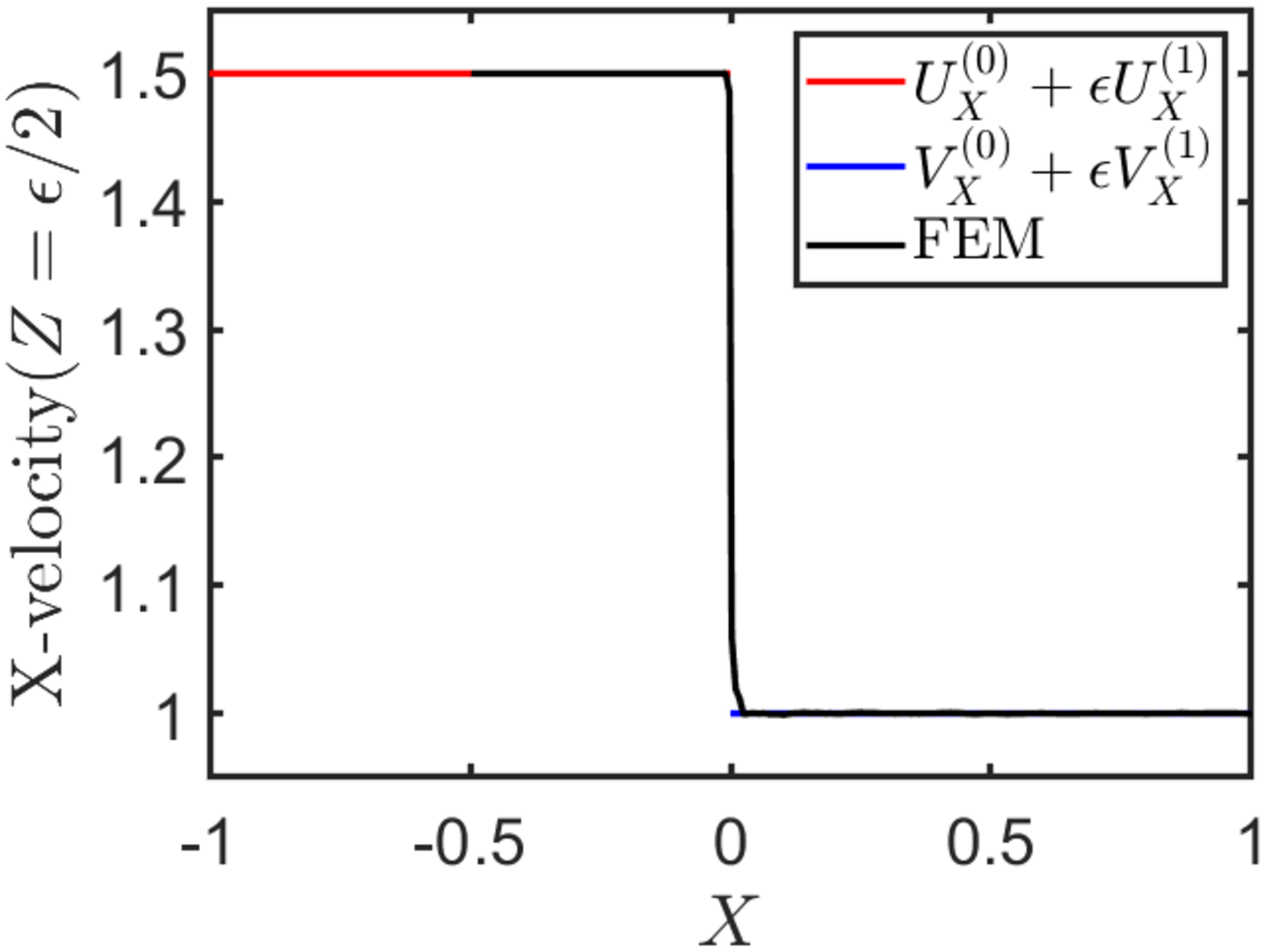}
			\includegraphics[width=0.325\textwidth]{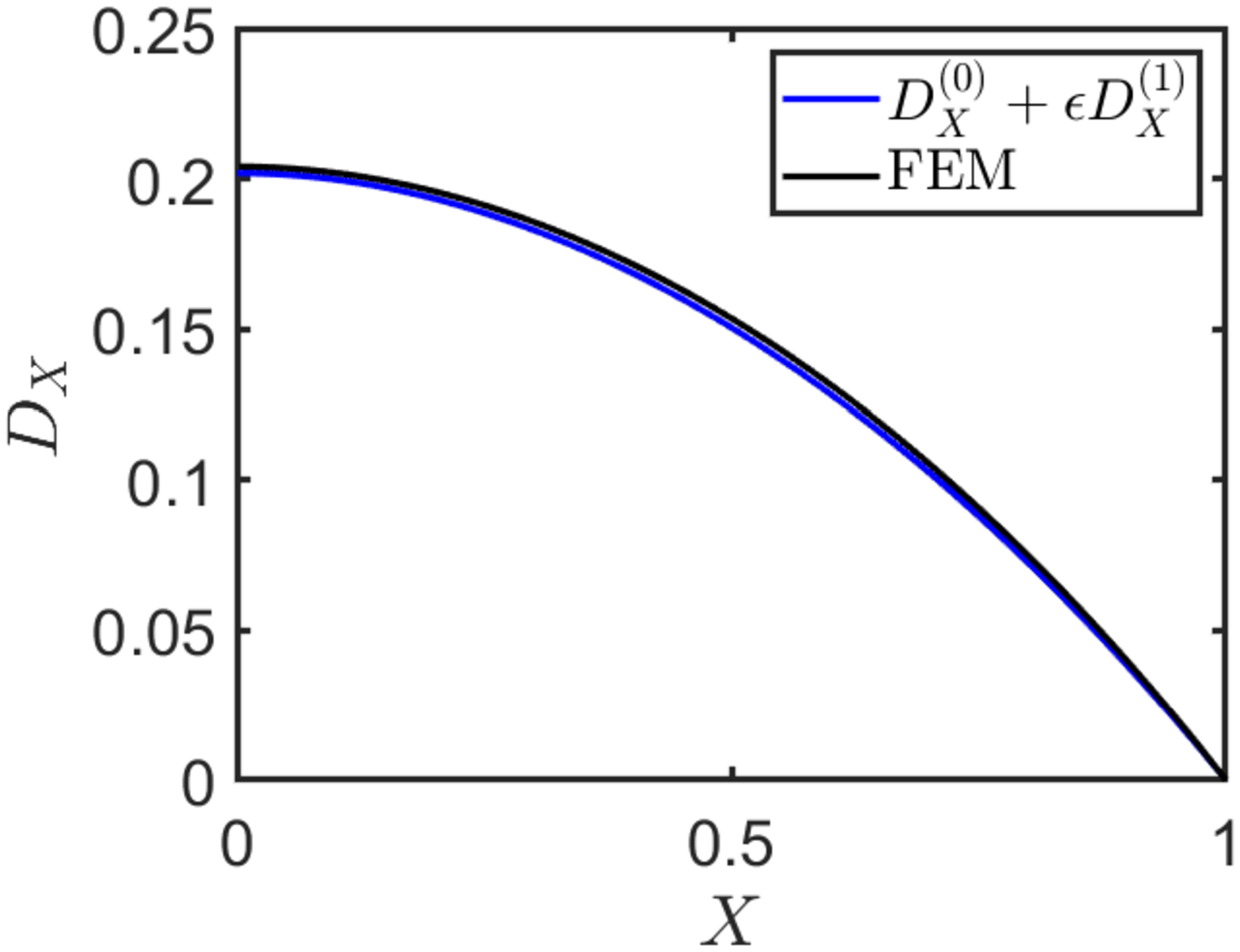}
		\end{center}
		
		\vspace{-4mm}
		\caption{Comparison between the lubrication limit and the finite element computations in terms of (left) pressure field, (centre) axial velocity field and (right) axial displacement field in the porous domain. Red and blue curves refer to the outer fluid and porous domains, respectively, and the black curves to finite element solutions. The agreement of the numerical solution is very good; to make the overlap with the analytical solution clearer, in the central panel we plot the black curve from the finite element results not covering the entire domain $X \in [-1,1]$.}\label{fig:Comparison_X}
	\end{figure}

	The problem solved with the finite element method presented in \S\ref{sec:fem} 
	is dimensional and we set the parameters values as $h = \epsilon$, $l = 1 $, $\mu_f = 1$, $\mu_{s,0} = 800000$,  $\gamma =0$ and $\kappa_0 = \bar{K} \epsilon^2 = 0.00015$, and having consistent units of measure. We consider flows driven by either a steady or a oscillatory upstream velocity profiles: in the latter case, we consider that the dimensional pulsation of velocity is equal to the inverse of the characteristic time.
	
	We start by presenting the steady case ($f(T)=1$).  In fig. \ref{fig:Comparison_X} we plot the pressure, axial velocity and axial displacement fields in the coupled Stokes flow-poroelastic domain, where the red and blue curves refer to quantities in the outer Stokes flow I and poroelastic domain IV, respectively, and the black curves to the finite element numerical results. All curves obtained within the lubrication limit are up to $O(\epsilon)$ except the pressure in the fluid domain that requires the solution of the inner problem in \eqref{EQ:StokesDarcy_Inner} (not solved here numerically). A good agreement (between analytical and numerical predictions) is observed. 
	
\begin{figure}[!t]
		\begin{center}
			\includegraphics[width=0.47\textwidth]{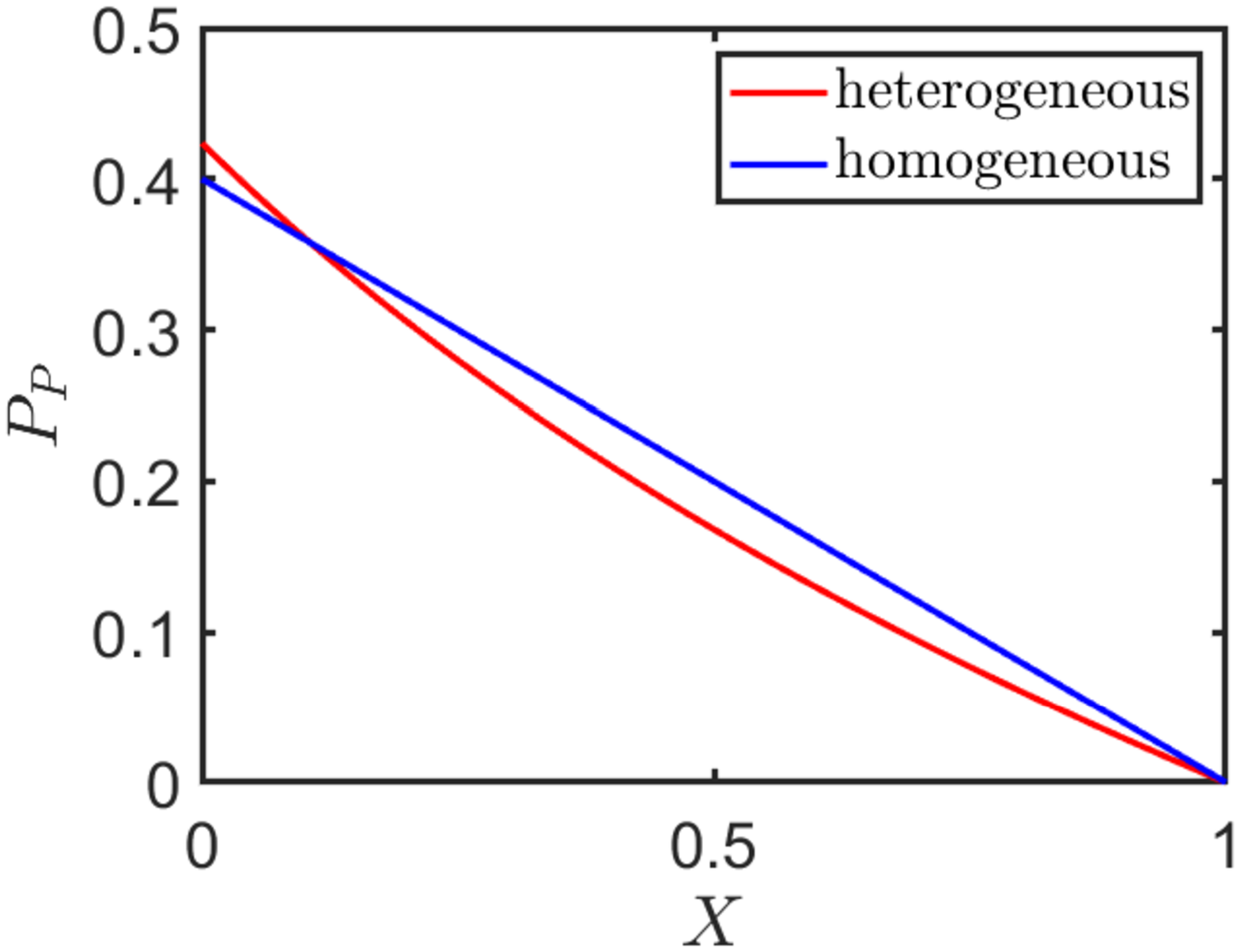}
			\includegraphics[width=0.47\textwidth]{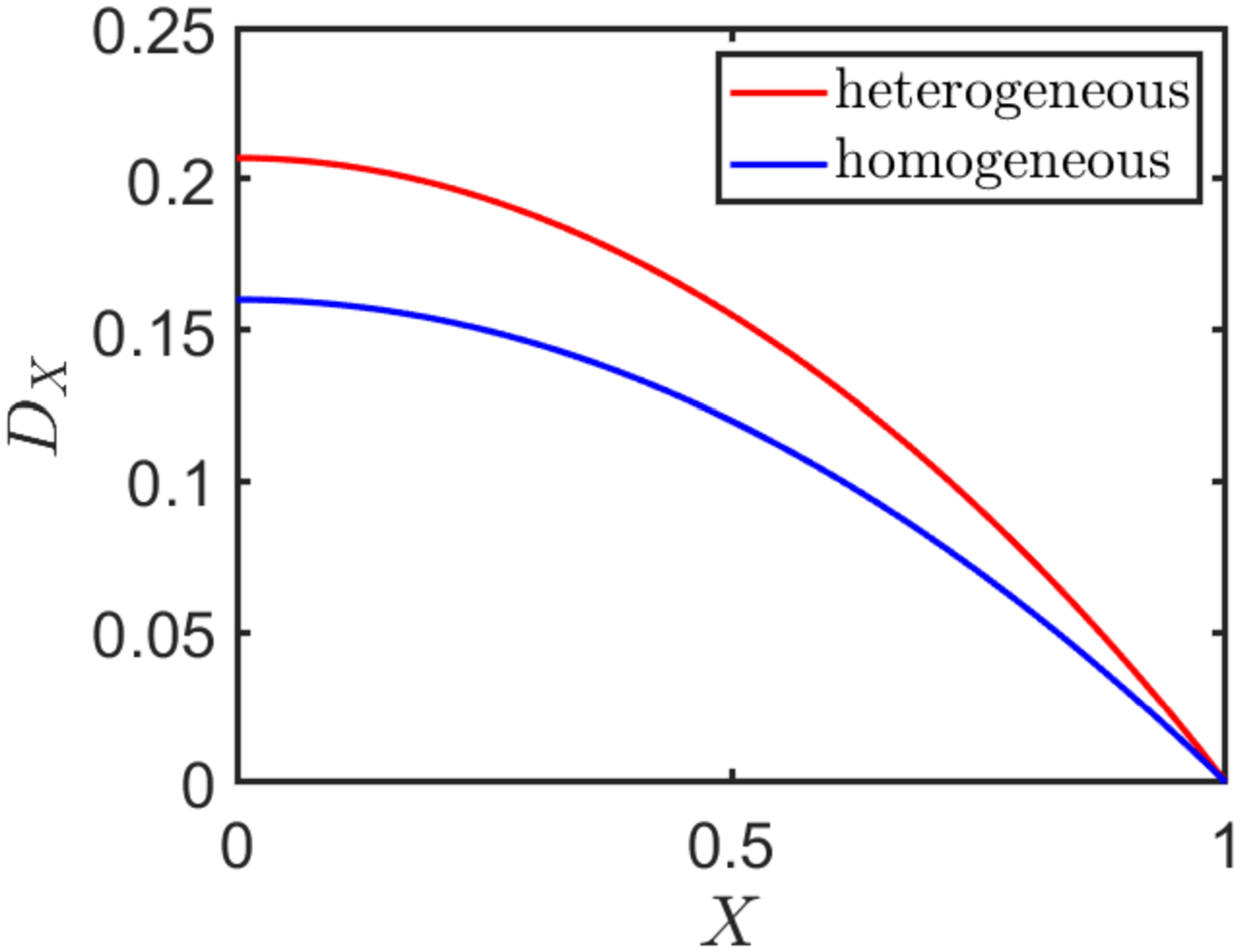}
		\end{center}
		
		\vspace{-4mm}
		\caption{Leading-order steady flow solution in term of pressure (left) and $X$-displacement (right) for the heterogeneous (red lines) and homogeneous (blue lines) case.}\label{fig:Static_IH_O1}
	\end{figure}

	Fig. \ref{fig:Static_IH_O1} shows the analytical pressure and axial displacement at leading-order in the poroelastic region, comparing homogeneous and  heterogeneous material properties. Although the spatial average of the material properties is the same at leading-order, the total pressure drop and the displacement of the interface $X=0$ differ. Moreover, if in the homogeneous case the pressure profile is necessarily linear and the displacement profile quadratic with respect to $X$, this is not true for the heterogeneous case: the choice of permeability field, that increases with $X$, leads to a concave pressure profile. A convex pressure profile is expected if the permeability decreases with $X$ as evident from the solutions \eqref{EQ:PP0_stat} and \eqref{EQ:PP1_stat}.
	
\begin{figure}[!t]
		\begin{center}
			\includegraphics[width=0.49\textwidth]{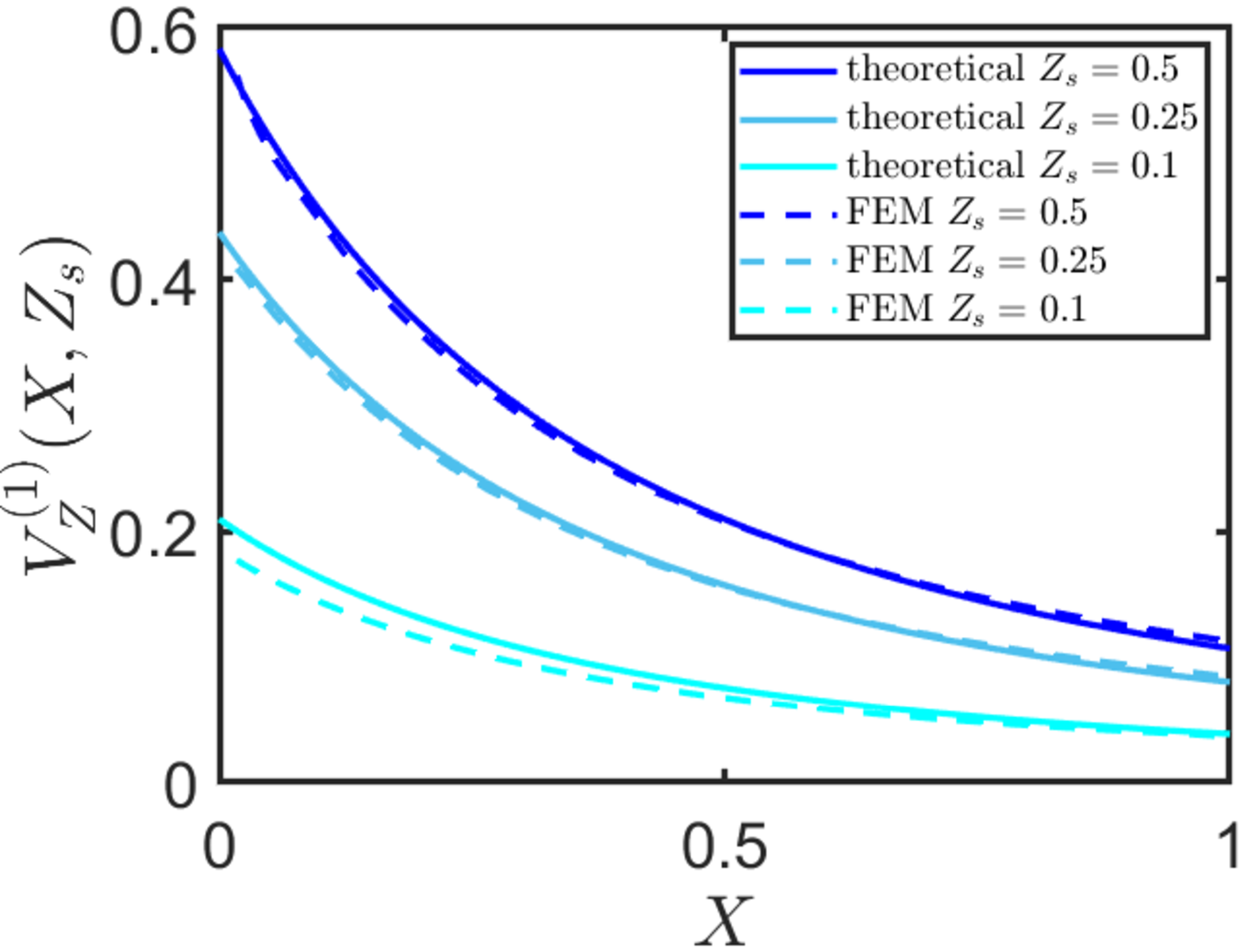}
			\includegraphics[width=0.49\textwidth]{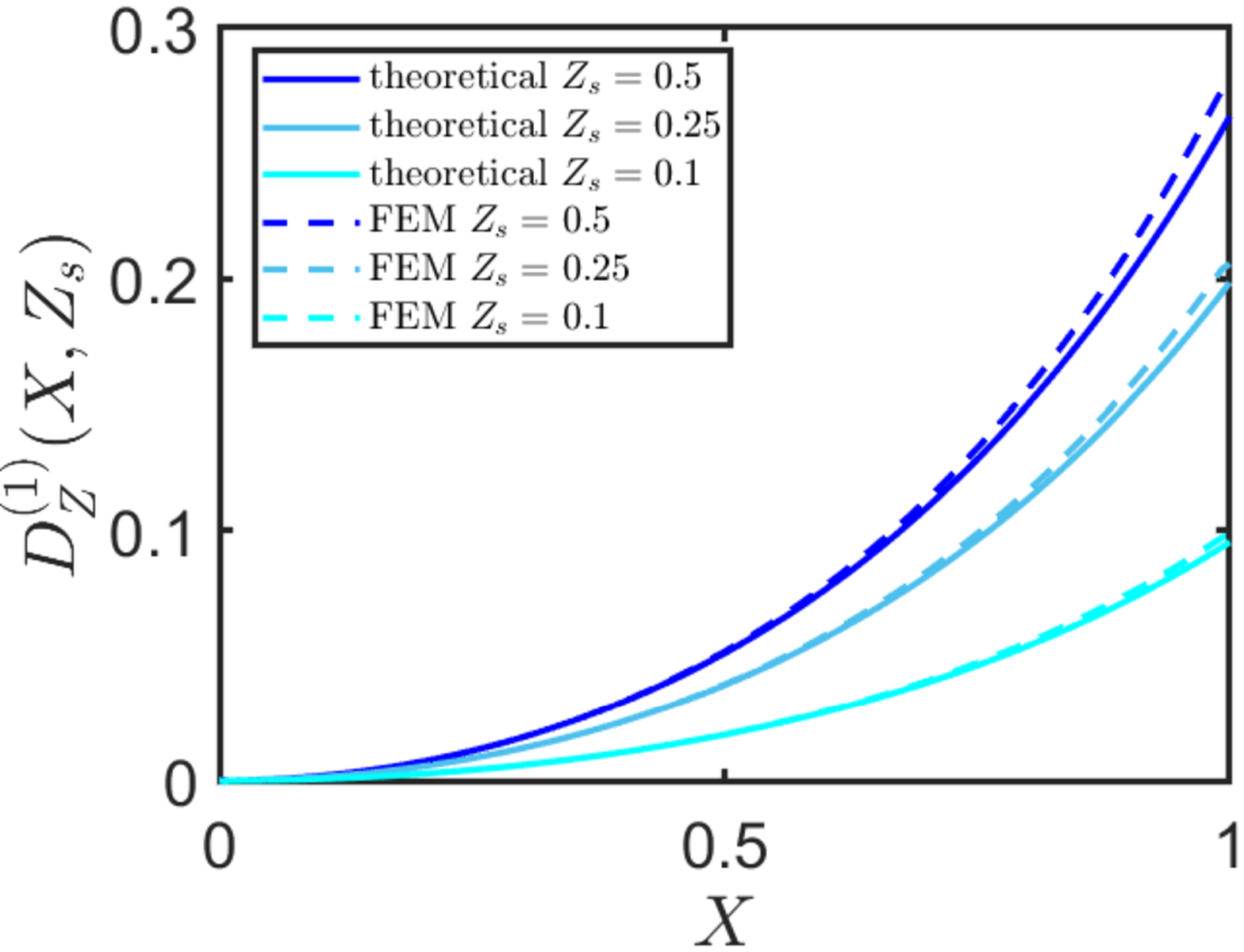}	
			\includegraphics[width=0.47\textwidth]{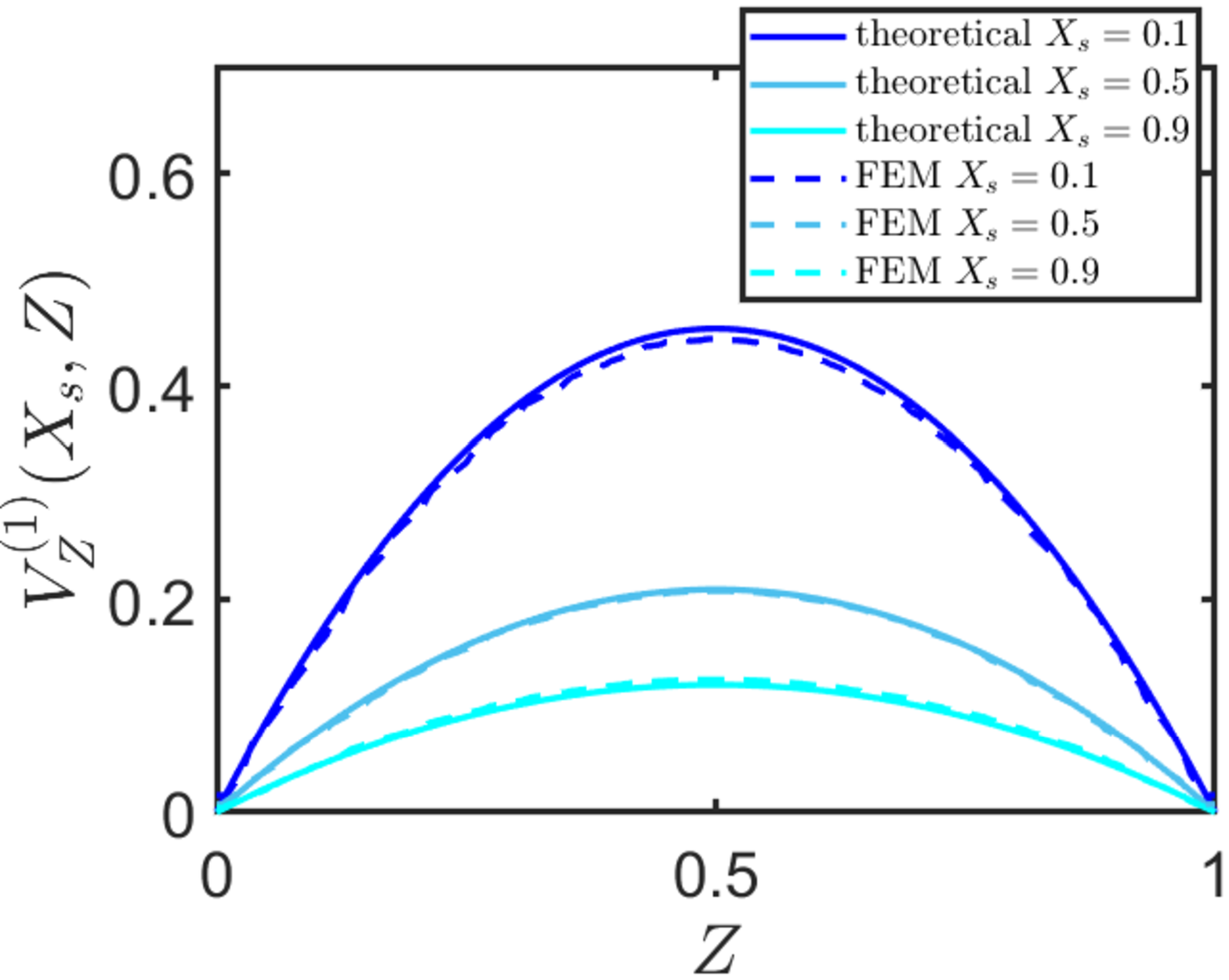}
			\includegraphics[width=0.47\textwidth]{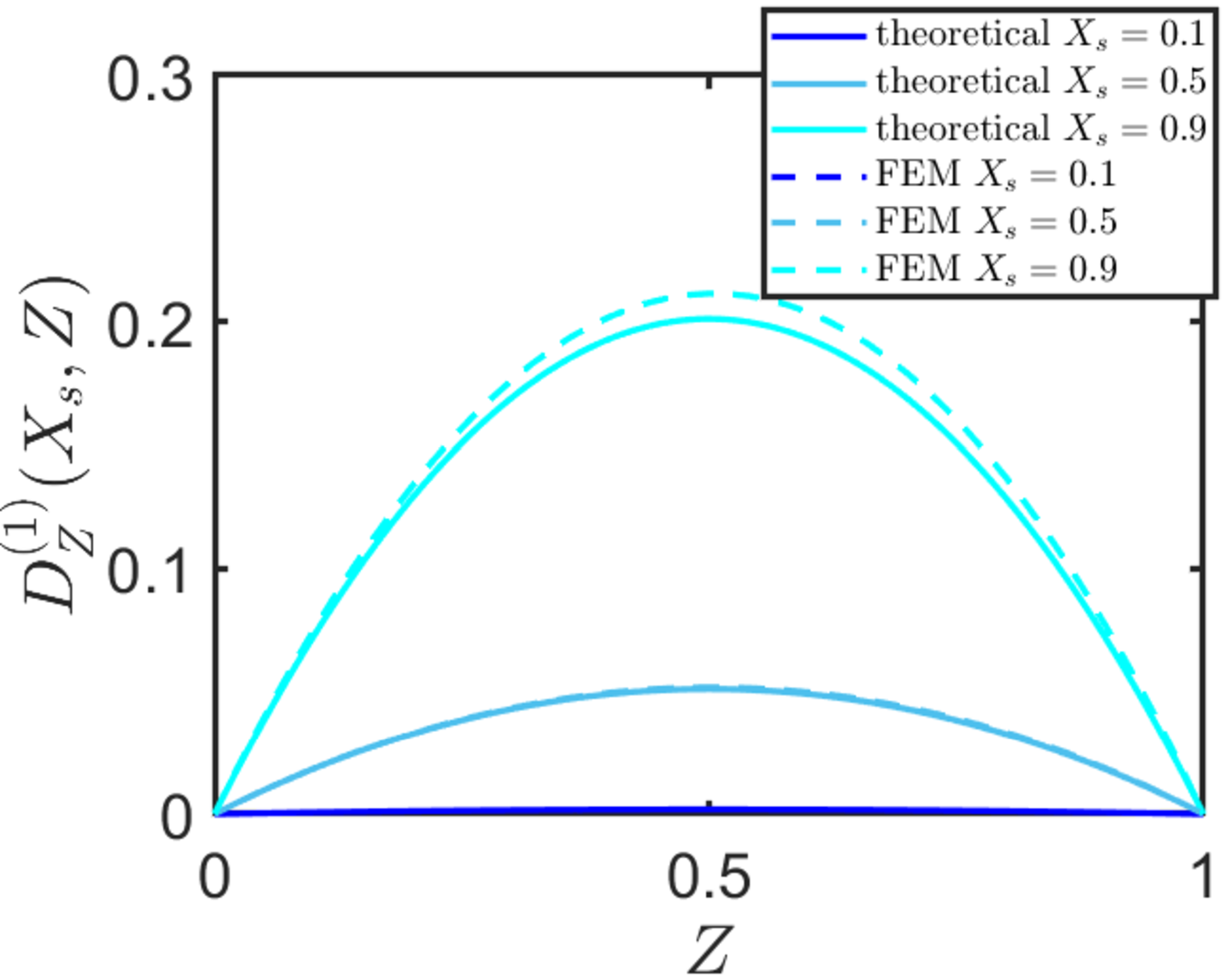}	
		\end{center}
		
		\vspace{-4mm}
		\caption{(top row) Comparison between the analytical (solid lines) and numerical (dashed lines) solutions for transverse Darcy velocity $V_Z^{(1)}(X,Z_s)$ (left) and transverse displacement  $D_Z^{(1)}(X,Z_s)$ (right) at fixed transverse coordinate. Transverse coordinates are indicated by dark blue ($Z_s=0.5$), mid blue ($Z_s=0.25$) and light blue ($Z_s=0.1$). (bottom row) Comparison between the analytical (solid lines) and numerical (dashed lines) solutions for transverse Darcy velocity $V_Z^{(1)}(X_s,Z)$ (left) and transverse displacement  $D_Z^{(1)}(X_s,Z)$ (right) at fixed longitudinal coordinate. Longitudinal coordinates are indicated by dark blue ($X_s=0.1$), mid blue ($X_s=0.5$) and light blue ($X_s=0.9$).}\label{fig:Comparison_Z}
	\end{figure}

	The heterogeneous distribution of the material properties causes the velocity and the displacement fields to depend on the transverse direction: the transverse Darcy velocity $V_{Z}$ and the transverse displacement $D_Z$ in domain IV are nonzero at $O(\epsilon)$ (see the analytical solutions \eqref{EQ:D1_stat} and \eqref{EQ:Vd1_stat}) and this is a key difference against the homogeneous case where $V^{(1)}_{Z}=0$ and $D_Z^{(1)}=0$. In fig. \ref{fig:Comparison_Z} (top row) we compare the analytical prediction and the numerical result for those two quantities at fixed selected transverse coordinates $Z_s = [0.1, 0.25, 0.5]$. In fig. \ref{fig:Comparison_Z} (bottom row), instead, we plot these quantities with respect to the coordinate $Z$ and at selected longitudinal coordinates $X_s=[0.1,0.5,0.9]$. With the material parameters as in \eqref{eq:choice}, all these profiles are parabolic in the transverse direction: however, the transverse velocity decreases with the longitudinal position whereas the transverse displacement increases moving towards the outlet. In the steady scenario, the fluid flow and the elastic deformation are always decoupled and thus we expect that the velocities and pressure profiles presented in both fig. \ref{fig:Static_IH_O1} and fig. \ref{fig:Comparison_Z} remain valid if we consider a rigid skeleton, with the displacement being obviously zero.
	
	The availability of closed form solutions allows us to easily investigate secondary variables such as the elastic stress distribution in the poroelastic domain. The components of the stress tensor are presented in Appendix \ref{APP:B}. In the top row of  fig. \ref{fig:Stresses_Lub}, we show the leading-order longitudinal $\Sigma_{XX}'^{(0)}$ and transverse $\Sigma'^{(0)}_{ZZ}$ stresses: they depend only on  $X$ but, in contrast to the homogeneous case, the increase (in magnitude) from inlet to outlet is not linear, because $D_X$ is no longer quadratic. The bottom row shows the $O(\epsilon)$ terms and they originate from the transverse heterogeneity of the material properties. For the specific case analysed we note that: (i) the longitudinal term is tensile and increasing moving towards the outlet and the top boundary; (ii) the transverse term is tensile, it depends only on $X$ and increases moving towards the outlet; (iii) the shear term, absent at the leading order, increases when moving from the inlet to the outlet and from the top/bottom boundaries to the centre; (iv) the components of the stress at the leading-order are positive because the permeability increases and the elastic parameters decreases moving from the inlet to the outlet.
	
	\begin{figure}[!t]
		\begin{center}
			\includegraphics[width=0.32\textwidth]{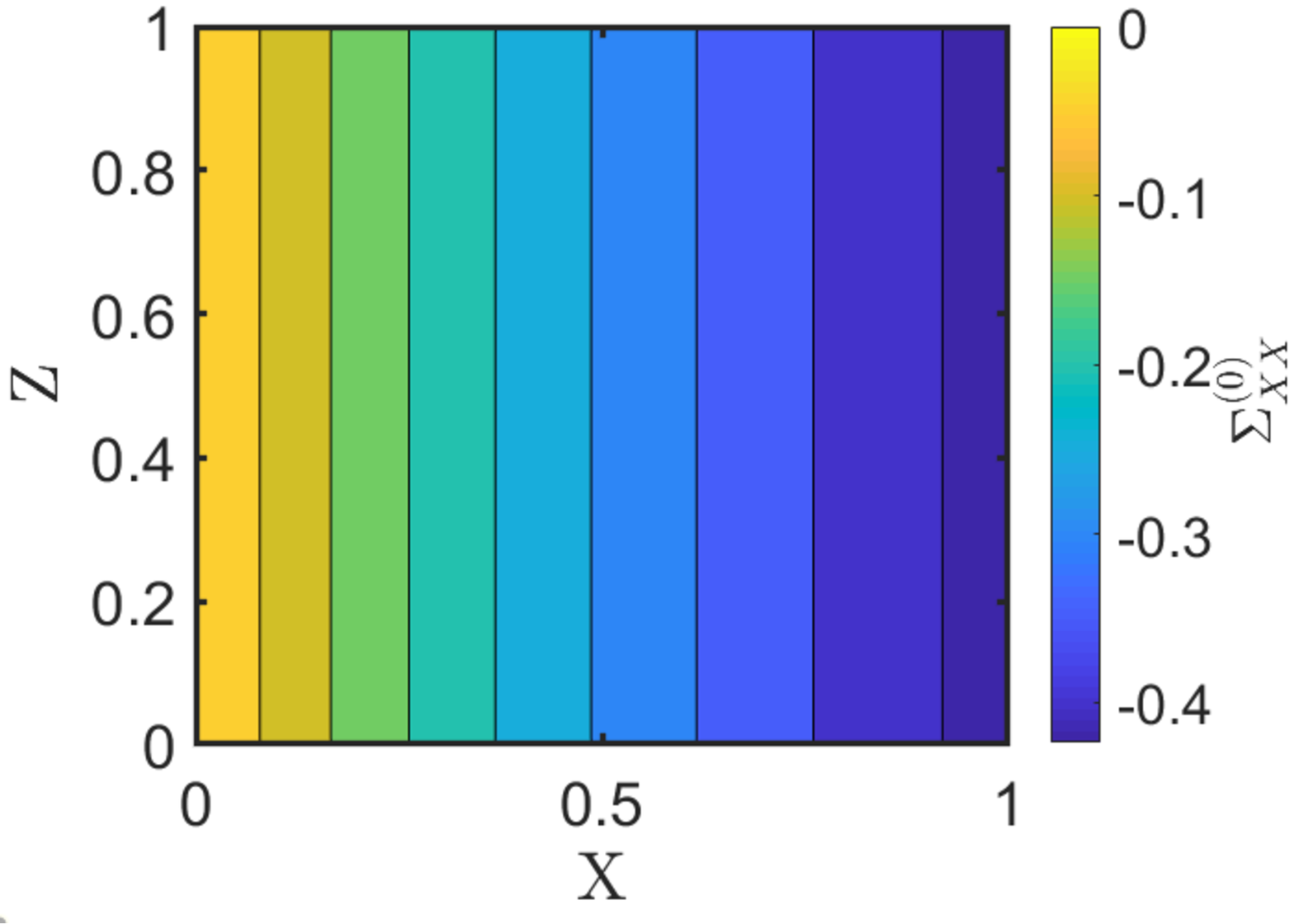}
			\includegraphics[width=0.32\textwidth]{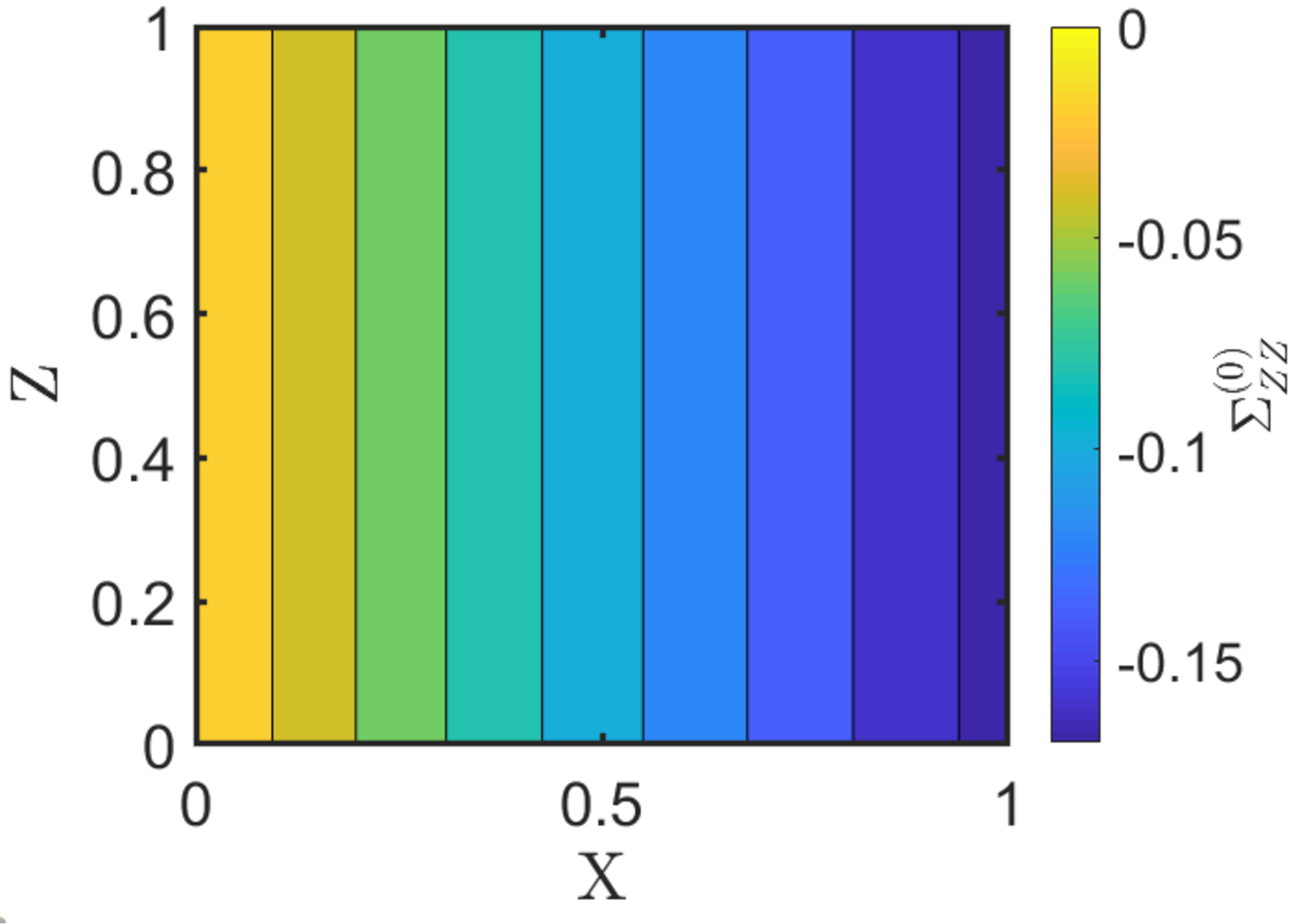}\\	
			\includegraphics[width=0.32\textwidth]{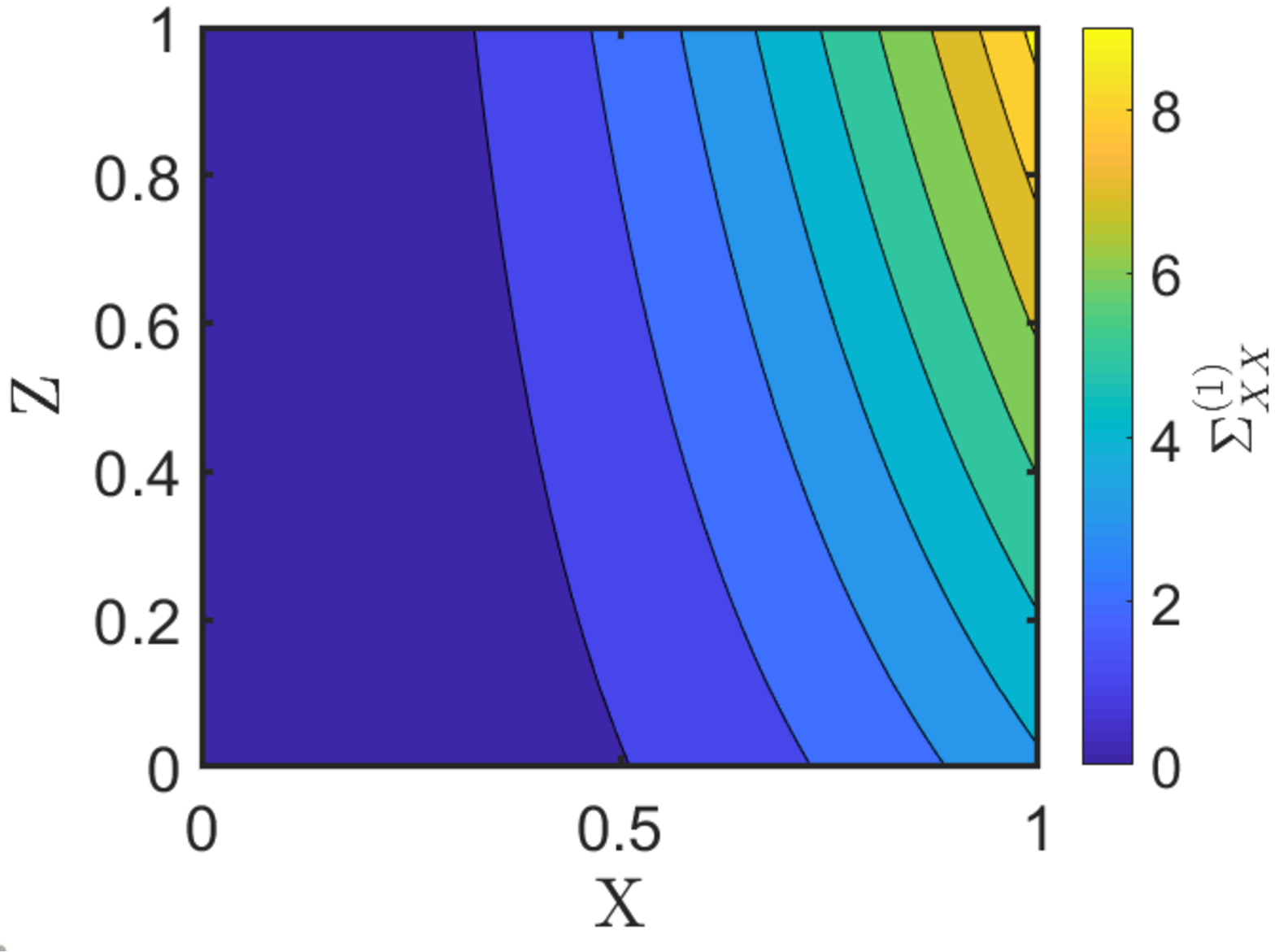}
			\includegraphics[width=0.32\textwidth]{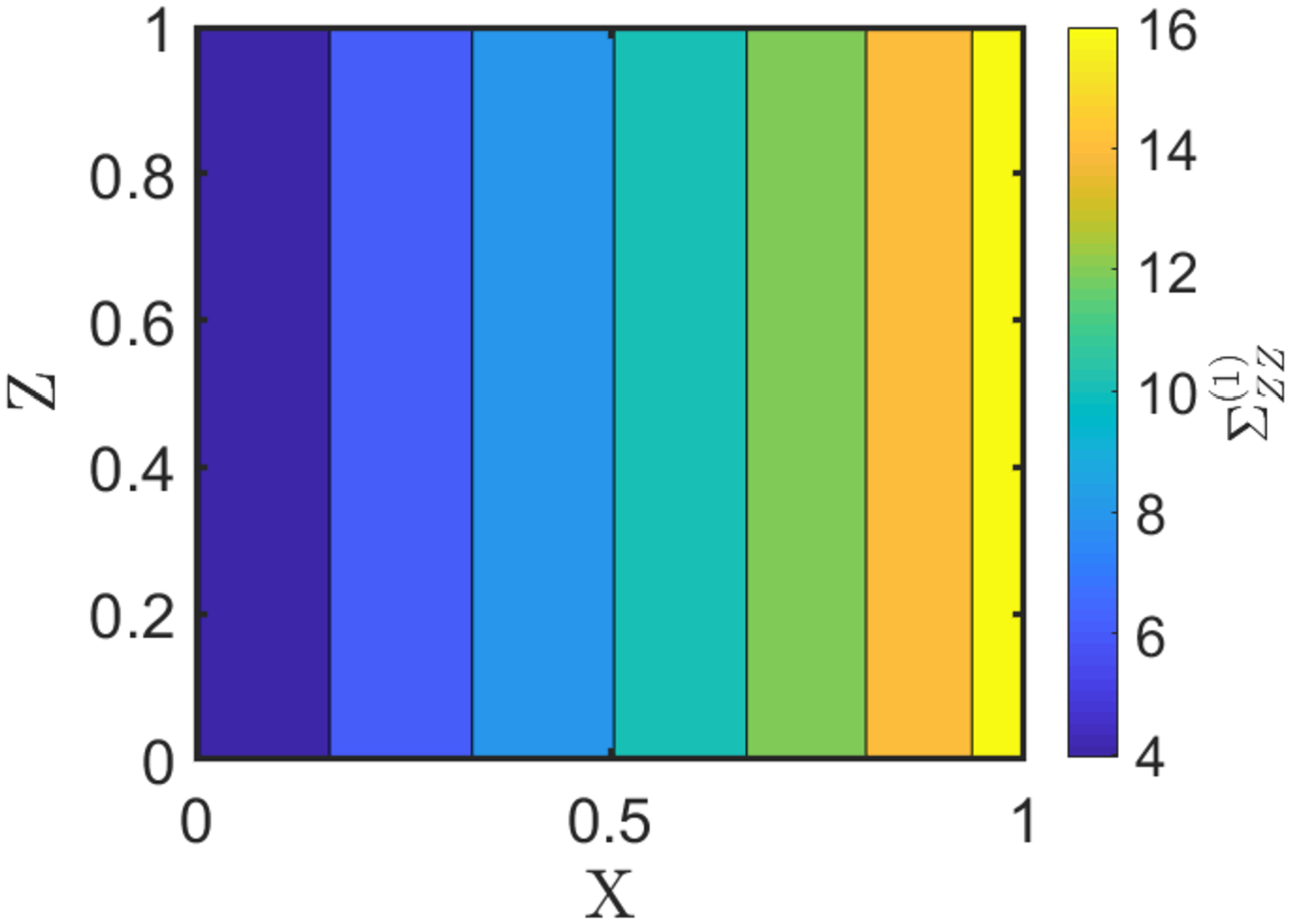}	\includegraphics[width=0.32\textwidth]{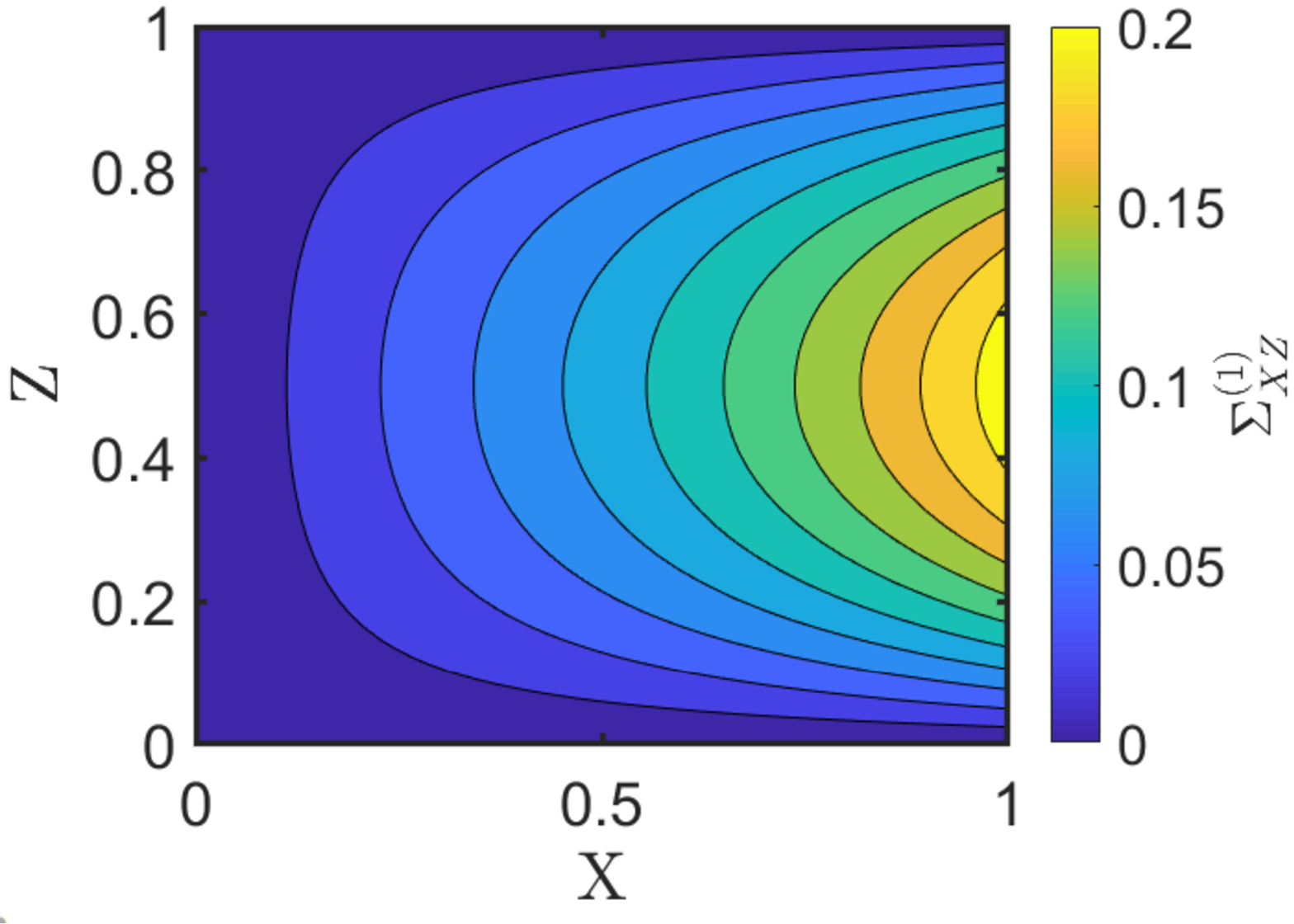}
		\end{center}
		
		\vspace{-4mm}
		\caption{(top row) Leading-order terms for elastic stress distribution in the poroelastic domain as $\Sigma'^{(0)}_{XX}$ (left) and $\Sigma'^{(0)}_{ZZ}$ (right). (bottom row) Order $O(\epsilon)$ terms for the elastic stress distribution in the poroelastic domain as $\Sigma'^{(1)}_{XX}$ (left), $\Sigma'^{(1)}_{ZZ}$ (center) and $\Sigma'^{(1)}_{XZ}$ (right). Note that colourbars differ for each plot.}\label{fig:Stresses_Lub}
	\end{figure}
	
\begin{figure}[!t]
		\begin{center}
			\includegraphics[width=0.47\textwidth]{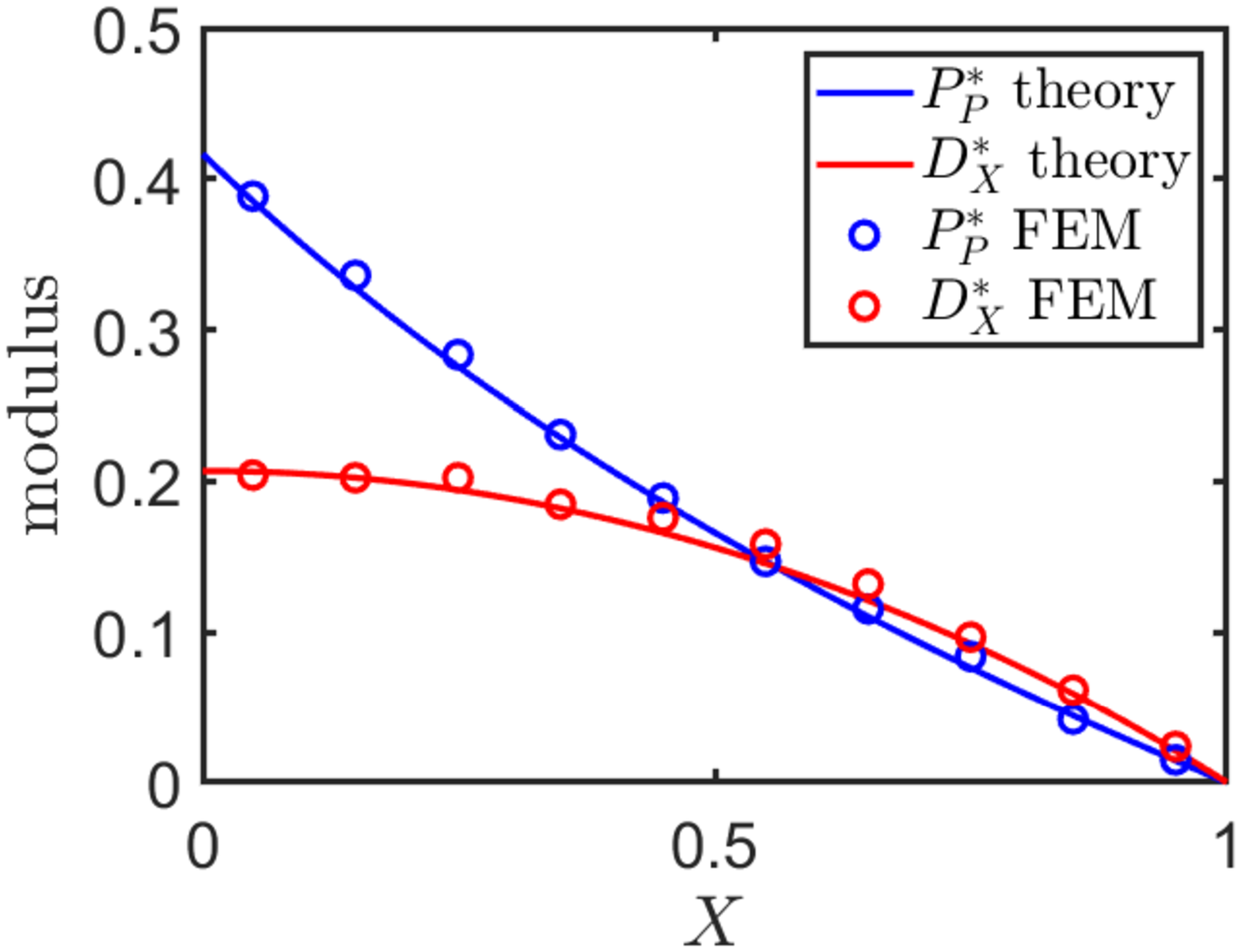}
			\includegraphics[width=0.47\textwidth]{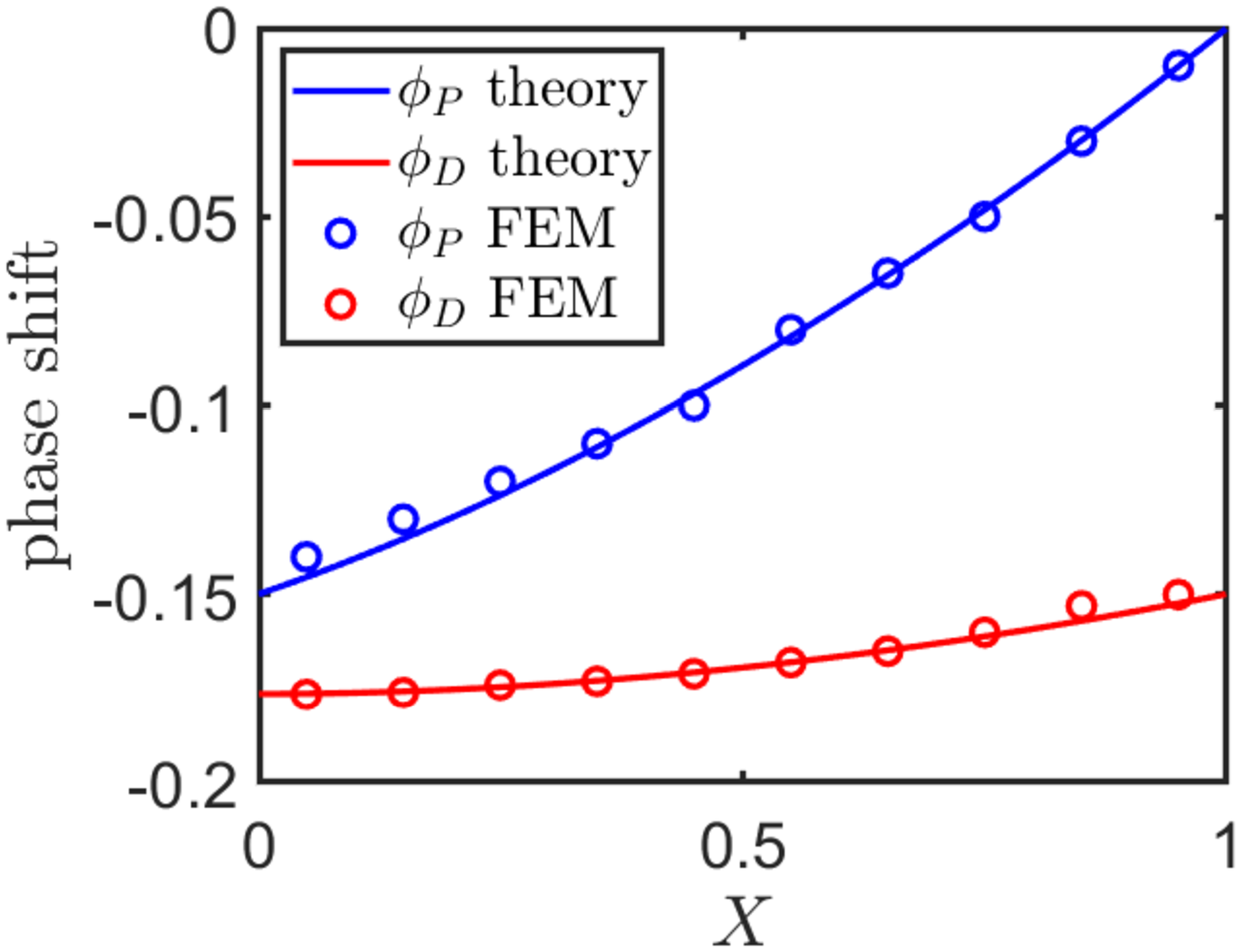}
		\end{center}
		
		\vspace{-4mm}
		\caption{Comparison between the oscillatory solution from the lubrication theory and the finite element solutions, in terms of modulus (left) and phase shift (right) of the pressure $P_P$ (blue) and the longitudinal displacement $D_X$ (red). Data shown here refer to centre-line $Z=0.5$. Curves are the theoretical solutions and markers are the results from the finite element simulations.}\label{fig:Oscillatory}
	\end{figure}

	In fig. \ref{fig:Oscillatory} we present the comparison in the oscillatory regime ($f(T)=\cos \left(T\right)$) in terms of the pressure  $P_P = P^*_P \cos( T+\phi_P)$ and the longitudinal displacement $D_X= D^*_X \cos( T+\phi_D)$ in domain IV. For both quantities we present the modulus and the phase shift with respect to the forcing $f(T)$ computed at the centre-line $Z=0.5$ and for different coordinates $X$: this information completely describes the time dependence  at any point $(X,Z=0.5)$ of the poroelastic domain; the continuous lines refer to the solutions in the lubrication limit and the empty circles to the finite element simulations. The $X$ dependence of the modulus is, as expected, the same as in the steady case (the material parameters are kept the same). Although we do not consider inertia, the negative phase shift shows that both the quantities are delayed with respect to $F(T)$ due to the fluid-elastic coupling present in domain IV. The phase shifts will be identically zero in the steady case or in the rigid case.
	
	\section{Concluding remarks}\label{sec:concl}
	In this work we investigate the mechanics of a single-phase fluid-poroelastic system, coupled via appropriate interfacial conditions. In the fluid domain we assume a Newtonian incompressible fluid while the linear poroelastic domain has an inhomogeneous porosity that induces heterogeneity of the permeability and the Lam\'e parameters, which we capture by prescribing the spatial dependence of the  material properties. 
	
	Focusing on the scenario where the characteristic time scale in the poroelastic domain ensures that the fluid and solid skeleton velocities are comparable, it can be shown that the coupling between a single-phase fluid domain and a poroelastic domain is controlled by two dimensionless parameters, $K_n$ and $U_n$. We apply the theoretical framework to a two-dimensional rectangular geometry, as typical, for example, in perfusion bioreactor systems. We exploit the small aspect ratio  $\epsilon=a/L\ll1$ of the channel and  perform an asymptotic analysis to determine closed-form analytical solutions to relate the spatial distribution of the material properties to the displacement, pressure, velocity and stress fields. (see \S\ref{sec:channel}). In this lubrication limit, there are four zones: the outer domains I (fluid) and IV (porous), where the mechanics depends on the interface via the coupling conditions derived in \S\ref{SECSEC:Transmission}; the inner domains II (fluid) and III (porous), where the coupling is controlled by the interface via the equations in \eqref{EQ:inter-lubrication_inner}. 
	
	In the outer domain I, we recover the analysis in \cite{Dalwadi2016} when, to drive the flow, we specify the upstream velocity profile. In the outer domain IV we derive the governing equations up to  order $O(\epsilon)$ and obtain  solutions for the displacement field, the Darcy velocity and the pressure field. Employing the interfacial conditions, we show that the solutions in the outer problem at  leading-order can be fully computed once the equation \eqref{EQ:Evolutionary_condition_O1} is solved. At  order $O(\epsilon)$, instead, to compute the full solution to the outer problem, we require both the solution of the equation \eqref{EQ:Evolutionary_condition_Oeps} and the function $P_F^{10}(T)$, which can be determined only when the inner Stokes-Darcy problem is solved. Equations \eqref{EQ:Evolutionary_condition_O1}  and \eqref{EQ:Evolutionary_condition_Oeps} are the coupling  conditions that we propose to couple domain I and domain IV, in terms of its solid effective elastic behaviour and thus close the problem.
	
	We use the solutions obtained from our analysis to discuss both the effect of inhomogeneity and deformability of the porous domain.
	
	In the steady case, the elastic deformation of the poroelastic domain is decoupled from the fluid flow. In such a regime we can solve the equations \eqref{EQ:Evolutionary_condition_O1} and \eqref{EQ:Evolutionary_condition_Oeps} in closed form and write the solution for the pressure field, the Darcy velocity and the displacement field up to order $O(\epsilon)$ in terms of the functions that describe the material heterogeneity. Compared to the homogeneous case, the presence of heterogeneous material properties permits to have leading-order profiles for the pressure and the displacement that are not necessarily linear and quadratic, respectively, and to have nonzero first order transverse Darcy velocity and displacement. Moreover, we show that heterogeneity can give rise to stress contributions not present for homogeneous materials. Indeed, the first order longitudinal and transverse components of the stress tensor are non necessarily linear in $X$, and the first order shear component of the stress tensor is nonzero. 
	
	The deformability of the solid phase modulates the temporal dynamics of the pressures and velocities via the extra contribution given by functions $\partial A_D^{(0)}/\partial T$ and $\partial A_D^{(1)}/\partial T$. This is shown in fig. \ref{fig:Oscillatory} for a particular set of parameters. For comparison, in the case of a porous domain with a rigid solid skeleton, all the fields would have the same temporal behaviour as the imposed forcing at the inlet $f(T)$.
	
	To investigate the behaviour in domains II and III we consider the effect of the interface as described by the equations in \eqref{EQ:inter-lubrication_inner}. The interesting outcome of our analysis is that, for the inner problem, the elastic and flow subproblems are decoupled, even in the transient case and up to order $O(\epsilon)$. We find that the stress tensors at both leading-order $\bar{\bSigma}'^{(0)}$ and first order $\bar{\bSigma}'^{(1)}$ are divergence free, with no traction applied at the interface. Thus  region III undergoes  a rigid body motion of a constant displacement fixed by  domain IV. As a consequence, the fluid flow within the pores is regulated via a coupling between a Stokes domain and a Darcy domain with constant permeability, which is a much easier problem to solve numerically compared to the original Biot-Stokes system. The finding that the elastic part of domain III is stress free and the displacement homogeneous is the argument that allows us to provide the equations \eqref{EQ:Evolutionary_condition_O1}, \eqref{EQ:Evolutionary_condition_Oeps}, \eqref{EQ:int_V0} and \eqref{EQ:int_V1} to close the problem in the outer domains.
	
	A possible application of our work is to perfusion systems, that are designed to guide the growth of a tissue by providing mechanical load to cells \cite{ElHaj1990}. In such systems, cells are seeded within a porous biomaterial scaffold, which is then cultured in a perfusion bioreactor which forces fluid through the biomaterial scaffold.  In this context, our analysis could provide ready-to-use relations to understand how the scaffold properties can be exploited to control prescribed mechanical load to the cells, and how growth (captured via a a change in porosity) could induce a change in mechanical stimuli.
	
	Natural extensions of the present work include the analysis of the nonlinear elasticity case where large deformations are present. Our analysis in the lubrication limit can be applied to situations where there is a clear separation between the transverse and the longitudinal scales; however, the effect of the interface can be investigated in a more general setting, studying how the description proposed here in the two-dimensional Cartesian setting can be generalised to 2D interfaces in a three dimensional framework.
	
	\noindent
	{\bf Acknowledgements.}
	MT is a member of the Gruppo Nazionale di Fisica Matematica (GNFM) of the Istituto Nazionale di Alta Matematica (INdAM). RRB has been partially supported by the Monash Mathematics Research Fund S05802-3951284. SLW gratefully acknowledges funding from the EPSRC (Healthcare Technologies Awards EP/P031218/1 \& EP/S003509/1) and the MRC (MR/T015489/1). 
	
	\begin{appendix}
		\section{Derivation of the solution in the outer fluid domain I}\label{APP:A}
		Using the expansion \eqref{EQ:expansion_LL} onto the governing equations in the fluid domain \eqref{eq:taoM}, we obtain
		\begin{align}
		\label{EQ:equilibriumF_X}
		\left(\frac{\partial U^{(0)}_{X}}{\partial Z^2} - \frac{\partial P_F^{(0)}}{\partial X} \right)+ \epsilon \left(\frac{\partial U^{(1)}_{X}}{\partial Z^2} - \frac{\partial P_F^{(1)}}{\partial X} \right) +  O(\epsilon^2) & = 0,\\ 
		\label{EQ:equilibriumF_Z}
		\left(\frac{\partial P_F^{(0)}}{\partial Z}\right) + \epsilon \left(\frac{\partial P_F^{(1)}}{\partial Z}\right)+  O(\epsilon^2) & = 0,\\
		\label{EQ:continuityF}
		\left(\frac{\partial U^{(0)}_{X}}{\partial X} + \frac{\partial U^{(0)}_{Z}}{\partial Z} \right)+ \epsilon \left(\frac{\partial U^{(1)}_{X}}{\partial X} + \frac{\partial U^{(1)}_{Z}}{\partial Z} \right) + O(\epsilon^2) & = 0.
		\end{align}
		At the leading-order, from \eqref{EQ:equilibriumF_Z} we obtain
		\begin{equation}
		P_F^{(0)} = P_F^{(0)} (X,T),
		\end{equation}
		and, from \eqref{EQ:equilibriumF_X},
		\begin{equation}
		U_{X}^{(0)} = \frac{Z^2}{2} \frac{\partial P_F^{(0)}}{\partial X} + Z U^{01}_{X} (X,T) + U^{00}_{X}(X,T),
		\end{equation}
		with $U^{00}_{X}(X,T) = 0$ and $U^{01}_{X} (X,T) = -\left(\partial P_F^{(0)}/\partial X\right)/2$ when imposing $U_{X}^{(0)} (Z =0,1)=0$ from the no-slip condition \eqref{EQ:BC_noslipF}.
		From \eqref{EQ:continuityF} we can thus obtain the leading velocity along $Z$ as
		\begin{equation}
		U_{Z}^{(0)} = U^{00}_{Z} (X,T)-\frac{1}{2} \left(\frac{Z^3}{3}-\frac{Z^2}{2}\right) \frac{\partial^2 P}{\partial X^2}.
		\end{equation}
		Using the no-slip condition $U_{Z}^{(0)}(Z=0) =0$, we get $ U^{00}_{Z} (X,T) = 0$ while from  $U_{Z}^{(0)}(Z=1) =0$ we have an equation for the pressure (decaying linearly with $X$), that in turn gives
		\begin{equation}
		U_{X}^{(0)} = \frac{1}{2}P_F^{01}\left(Z^2-Z\right).
		\end{equation}
		At the leading-order, the condition on the imposed velocity at the inlet \eqref{EQ:BC_inlet} reduces to  
		\begin{equation}
		\int_{0}^{1} U_{X}^{(0)} \upd Z = f(T),
		\end{equation}
		and we get $P_F^{01}(T)=-12 f(T)$.
		At  $O(\epsilon)$, the governing equations are formally equal to what we got at the leading-order once using the second term in the expansion for the variables. Thus 
		\begin{equation}
		P_F^{(1)} = P_F^{10}(T) + P_F^{11}(T) X, \qquad 
		U_{X}^{(1)} = \frac{1}{2}P_F^{11}(T)\left(Z^2-Z\right), \qquad 
		U_{Z}^{(1)} = 0,
		\end{equation} 
		with the condition on the imposed velocity at the inlet \eqref{EQ:BC_inlet} at the first order to be
		\begin{equation}
		\epsilon^2 \int_{0}^{1} U_{X}^{(1)} \upd Z = 0.
		\end{equation}
		From this  we obtain $P_F^{11}(T) = 0$, thus recovering the expressions \eqref{EQ:Fluid_Outer0} and \eqref{EQ:Fluid_Outer1}.
		
		\section{Derivation of the solution in the outer poroelastic domain IV}\label{APP:B}
		The leading-order problem in domain IV reads
		\begin{subequations}
			\begin{align}
			\label{EQ:equilibrium_IV_O1}
			&
			\frac{\partial \Sigma_{XZ}'^{(0)} }{\partial Z} = 0,\ 
			\frac{\partial \Sigma_{XZ}'^{(0)} }{\partial X}  + \frac{\partial}{\partial Z}\left(\Sigma_{ZZ}'^{(0)}- P_P^{(0)}\right)=0, \tag{\theequation a,b}
			\\
			\label{EQ:darcy_IV_O1}
			&
			V_{X}^{(0)} = - \cK_0^{(0)}\left(X\right)\displaystyle\frac{\partial P_P^{(0)}}{\partial X},\ 
			0 = - \displaystyle\frac{\partial P_P^{(0)}}{\partial Z}, \tag{\theequation c,d}\\
			\label{EQ:continuity_IV_O1}
			&\frac{\partial}{\partial X} \left(\frac{\partial D_X^{(0)}}{\partial T} +  V_{X}^{(0)} \right) + \frac{\partial}{\partial Z} \left(\frac{\partial D_Z^{(0)}}{\partial T}  + V_{Z}^{(0)}\right) =0, \tag{\theequation e}\\
			\label{EQ:StressDisp_IV_O1}
			&\Sigma_{XZ}'^{(0)} = \cH_0^{(0)}\frac{\partial D_X^{(0)}}{\partial Z}, \quad \Sigma_{ZZ}'^{(0)} = \left(2 \cH_0^{(0)}+\cN_0^{(0)}\right)\frac{\partial D_Z^{(0)}}{\partial Z} + \cN_0^{(0)}\frac{\partial D_X^{(0)}}{\partial X},\tag{\theequation f}
			\end{align}\end{subequations}
		together with boundary conditions
		\begin{subequations}
			\label{plantlead}
			\begin{align}
			\label{EQ:BC_noslipPlead}
			(D^{(0)}_Z, V^{(0)}_{Z},\Sigma_{XZ}'^{(0)})=(0,0,0)  \quad \text{on} \quad Z=0,1 & \quad \text{for} \quad X>0,\\
			\label{EQ:BC_outletlead}
			D^{(0)}_X = 0,\quad P_P^{(0)} = 0 & \quad \text{at} \quad X = 1.
			\end{align}\end{subequations}
		
		From equation \eqref{EQ:darcy_IV_O1} we obtain 
		\begin{equation}\label{EQ:PVx_OuterFull0}
		P_P^{(0)} = P_P^{(0)}(X,T), \quad V_{X}^{(0)} = -  \cK_0^{(0)} \frac{\partial P_P^{(0)}}{\partial X}. \tag{\theequation a,b}
		\end{equation}
		Equation \eqref{EQ:continuity_IV_O1} subject to the condition $V^{(0)}_{Z}(X,Z=0,T)=0$ gives
		\begin{equation}\label{EQ:Vz_OuterFull0}
		V_{Z}^{(0)}=- \int_0^Z \frac{\partial }{\partial T}\left(\nabla \cdot \bD^{(0)}\right) \upd \zeta + Z \frac{\partial }{\partial X} \left(\cK_0^{(0)}\frac{\partial P_P^{(0)}}{\partial X} \right).
		\end{equation}
		Exploiting the boundary condition $V^{(0)}_{Z} (X,Z=1,T) =0$, together with the boundary condition $P_P^{(0)}(X=1)=0$  we obtain the following expression for the pressure
		\begin{equation}\label{EQ:P_OuterFull0}
		P_P^{(0)} = P_P^{01}(T)\int_1^X \frac{1}{\cK_0^{(0)}}\upd \xi + \int_1^X  \frac{1}{\cK_0^{(0)}}\left(\int_{0}^{\xi}\left(\int_0^1  \frac{\partial }{\partial T}\left(\nabla \cdot \bD^{(0)}\right)\upd \zeta\right)\upd \eta\right)\upd \xi.
		\end{equation}
		The displacement and stress  are computed from \eqref{EQ:equilibrium_IV_O1},\eqref{EQ:StressDisp_IV_O1}, using the boundary conditions $D_Z^{(0)}(X,Z=0,T)=D_Z^{(0)}(X,Z=1,T)=0$ and $\Sigma_{XZ}'^{(0)}(X,Z=0,T) = \Sigma_{XZ}'^{(0)}(X,Z=1,T)= 0$. We find
		\begin{equation}
		\label{EQ:DX_OuterFull0}
		\bD^{(0)} = (A^{(0)}_D(X,T),0), \ \Sigma'^{(0)}_{XZ}=0, \ \Sigma'^{(0)}_{ZZ}=\cN_0^{(0)}\partial A^{(0)}_D/\partial X. \tag{\theequation a,b,c}
		\end{equation}
		The  $O(\epsilon)$-problem in domain IV reads
		\begin{subequations}
			\begin{align}
			\label{EQ:equilibrium_IV_Oeps}
			&
			\displaystyle\frac{\partial \Sigma_{XZ}'^{(1)} }{\partial Z} = 0,\ 
			\displaystyle\frac{\partial \Sigma_{ZZ}'^{(1)} }{\partial Z} - \frac{\partial P_P^{(1)}}{\partial Z}+ \frac{\partial \Sigma_{XZ}'^{(1)}}{\partial X}=0, \tag{\theequation a,b}
			\\
			\label{EQ:darcy_IV_Oeps}
			&
			\displaystyle V_{X}^{(1)} = - \left(\cK_0^{(0)}\displaystyle\frac{\partial P_P^{(1)}}{\partial X} + \cK_0^{(1)}\displaystyle\frac{\partial P_P^{(0)}}{\partial X} \right),\ 
			\displaystyle 0 = -\displaystyle\frac{\partial P_P^{(1)}}{\partial Z}, \tag{\theequation c,d}\\
			\label{EQ:continuity_IV_Oeps}
			&\frac{\partial}{\partial X} \left(\frac{\partial D_X^{(1)}}{\partial T}  + V_{X}^{(1)} \right) + \frac{\partial}{\partial Z} \left(\frac{\partial D_Z^{(1)}}{\partial T}   + V_{Z}^{(1)}\right) =0,\tag{\theequation e}
			\end{align}\end{subequations}
		with
		\begin{align}\label{EQ:StressDisp_IV_Oeps}
		\Sigma_{XZ}'^{(1)} &= \cH_0^{(0)}\frac{\partial D_X^{(1)}}{\partial Z}+\cH_0^{(1)}\frac{\partial D_X^{(0)}}{\partial Z}, \\
		\Sigma_{ZZ}'^{(1)} &= \left(2 \cH_0^{(0)}+\cN_0^{(0)}\right)\frac{\partial D_Z^{(1)}}{\partial Z} + \cN_0^{(0)}\frac{\partial D_X^{(1)}}{\partial X} + \cN_0^{(1)}\frac{\partial D_X^{(0)}}{\partial X},
		\end{align}
		together with boundary conditions
		\begin{subequations}
			\label{plantlead2}
			\begin{align}
			\label{EQ:BC_noslipP_Oeps}
			(D^{(1)}_Z, V^{(1)}_{Z},\Sigma_{XZ}'^{(1)})=(0,0,0)  \quad \text{on} \quad Z=0,1 & \quad \text{for} \quad X>0,\\
			\label{EQ:BC_outlet_Oeps}
			D^{(1)}_X = 0,\quad P_P^{(1)} = 0 & \quad \text{at} \quad X = 1.
			\end{align}\end{subequations}
		
		From equations \eqref{EQ:darcy_IV_Oeps} we obtain 
		\begin{equation}\label{EQ:PVx_OuterFull1}
		P_P^{(1)} = P_P^{(1)}(X,T), \quad V_{X}^{(1)} = -  \left(\cK_0^{(0)} \frac{\partial P_P^{(1)}}{\partial X}   + \cK_0^{(1)} \frac{\partial P_P^{(0)}}{\partial X}  \right). \tag{\theequation a,b}
		\end{equation}
		Equation \eqref{EQ:continuity_IV_Oeps} subject to the condition $V^{(1)}_{Z}(X,Z=0,T)=0$ gives
		\begin{equation}\label{EQ:Vz_OuterFull1}
		V_{Z}^{(1)}=-\int_0^Z \frac{\partial}{\partial T} \left( \nabla \cdot \bD^{(1)}\right) \upd \zeta + \int_0^Z \frac{\partial}{\partial X}  \left(\cK_0^{(1)} \frac{\partial P_P^{(0)}}{\partial X}\right) \upd \zeta + Z \frac{\partial}{\partial X}  \left(\cK_0^{(0)} \frac{\partial P_P^{(1)}}{\partial X}\right).
		\end{equation}
		Exploiting the boundary condition $V^{(1)}_{Z} (X,Z=1,T) =0$, together with the boundary condition $P_P^{(1)}(X=1)=0$  we obtain the following expression for the pressure
		\begin{equation}\label{EQ:P_OuterFull1}
		P_P^{(1)} = P_P^{11}(T)\int_1^X \frac{1}{\cK_0^{(0)}}\upd \xi + \int_1^X  \frac{1}{\cK_0^{(0)}}\left(\int_0^{\xi}\left(G^{(1)}(\eta,T) \right)\upd \eta\right)\upd \xi,
		\end{equation}
		with
		\begin{equation*}
		G^{(1)}= \int_0^1 \frac{\partial}{\partial T} \left(\nabla \cdot \bD^{(1)} \right)\upd \zeta - \int_0^1 \frac{\partial}{\partial X} \left(\cK_0^{(1)}\frac{\partial P_P^{(0)}}{\partial X}\right)\upd \zeta.
		\end{equation*}
		The displacement and stress fields are computed from \eqref{EQ:equilibrium_IV_Oeps} and  \eqref{EQ:StressDisp_IV_O1}, using the boundary conditions $D_Z^{(1)}(X,Z=0,T)=D_Z^{(1)}(X,Z=1,T)=0$ and $\Sigma_{XZ}'^{(1)}(X,Z=0,T) = \Sigma_{XZ}'^{(1)}(X,Z=1,T)= 0$. We find
		\begin{align}
		\label{EQ:DX_OuterFull1}
		\bD^{(1)} = \left(A^{(1)}_D(X,T),\frac{1}{2 \cH_0^{(0)}+\cN_0^{(0)}}\left( Z\int_0^1\cN_0^{(1)}\upd \zeta-\int_0^Z \cN_0^{(1)}\upd \zeta\right)\frac{\partial A_D^{(0)}}{\partial X}\right),\nonumber \\
		\Sigma'^{(1)}_{XZ}=0, \ \Sigma'^{(1)}_{ZZ}=\cN_0^{(0)}\partial A_D^{(1)}/\partial X + \cN_0^{(1)}\partial A_D^{(0)}/\partial X. \tag{\theequation a,b,c}
		\end{align}
		
	\end{appendix}


\begin{thebibliography}{0}
	
\bibitem{ElHaj1990}
A.J. El-Haj, S.L. Minter, S.C. Rawlinson, R. Suswillo and L.E. Lanyon, Cellular responses to mechanical loading in vitro, {\it J. Bone and Min. Res.} {\bf 5} (1990) 923 - 32.

\bibitem{Thomen2017}
P. Thomen, J. Robert, A. Monmeyran, A.F. Bitbol, C. Douarche and N. Henry,	Bacterial biofilm under flow: First a physical struggle to stay, then a matter of breathing, {\it PLoS ONE} {\bf 12} (2017) e0175197.

\bibitem{ambar19}
I. Ambartsumyan, E. Khattatov, T. Nguyen and I. Yotov, Flow and transport in fractured poroelastic media, {\it GEM Int. J. Geomath.} {\bf 10} (2019) 11.	

\bibitem{coussy2004}
O. Coussy, {\it Poromechanics}. (John Wiley \& Sons Ltd, Chichester, UK 2004).

\bibitem{cowin2007}
S.C. Cowin and S.B. Doty, {\it Tissue Mechanics}. (Springer-Verlag, New York, USA 2007).	

\bibitem{terzaghi}
K. Terzaghi, Principles of Soil Mechanics: I—Phenomena of Cohesion of Clays, {\it Engineering News-Record} {\bf 95} (1925) 742 - 746.

\bibitem{biot41}
M.A. Biot,	General theory of three-dimensional consolidation, {\it J. Appl. Phys.} {\bf 12} (1941) 155 - 164.

\bibitem{lang16}
G.E. Lang, D. Vella, S.L. Waters and A. Goriely, Mathematical modelling of blood-brain barrier failure and oedema, {\it Math. Med. Biol.} {\bf 34} (2016) 391 - 414.

\bibitem{ateshian10}
G.A. Ateshian and J.A. Weiss, Anisotropic hydraulic permeability under finite deformation, {\it J. Biomech. Eng.} {\bf 132} (2010) 111004(7).

\bibitem{federico07}
S. Federico and W. Herzog, On the anisotropy and inhomogeneity of permeability in articular cartilage, {\it Biomech. Model. Mechanobiol.} {\bf 7} (2007) 367 - 378.

\bibitem{nedjar13}
B. Nedjar, Formulation of a nonlinear porosity law for fully saturated porous media at finite strains, {\it J. Mech. Phys. Solids} {\bf 61} (2013) 537 - 556.

\bibitem{macminn2016}
C.W. MacMinn, E.R Dufresne and J.S Wettlaufer, Large deformations of a soft porous material, {\it Phys. Rev. Applied} {\bf 5} (2016) 044020(30).

\bibitem{Mei1997}
C.C. Mei, J.I Auriault and F.J. Ursell, Mechanics of heterogeneous porous media with several spatial scales, {\it Proc. Royal Soc. London A} {\bf 426} (1997) 391 - 423.

\bibitem{Saez1989}
A.E. Sáez, C.J Otero and I. Rusinek, The effective homogeneous behavior of heterogeneous porous media, {\it Transport Porous Med.} {\bf 4} (1989) 213 - 238.

\bibitem{showalter05}
R.E. Showalter, Poroelastic Filtration Coupled to Stokes Flow, pp. 229 - 241 , in O. Imanuvilov, G. Leugering, R. Triggiani and B.Y. Zhang, {\it Control Theory of Partial Differential Equations - Lecture Notes in Pure and Applied Mathematics vol. 242} (Chapman \& Hall, Boca Raton, 2005).

\bibitem{Dalwadi2016}
M. Dalwadi, S.J. Chapman, S.L. Waters and J. Oliver, On the boundary layer structure near a highly permeable porous interface, {\it J. Fluid Mech.} {\bf 798} (2016) 88 - 139.

\bibitem{ruiz21}
R. Ruiz-Baier, M. Taffetani, H.D. Westermeyer, and I. Yotov, The Biot-Stokes coupling using total pressure: formulation, analysis and application to interfacial flow in the eye, {\it Comput. Methods Appl. Mech. Eng.} under revision (2021). Preprint available from http://arxiv.org/abs/2106.09191

\bibitem{Preziosi2002}
L. Preziosi and A. Farina, On Darcy's law for growing porous media, {\it Int. J. Nonlin. Mech.} {\bf 37} (2002) 485 - 491.

\bibitem{serpilli19}
M. Serpilli, Classical and higher order interface conditions in poroelasticity, {\it Ann. Solid Struct. Mech.} {\bf 11} (2019) 1 - 10.

\bibitem{murad01}
M.A. Murad, J.N. Guerreiro and A.F.D. Loula, Micromechanical computational modeling of secondary consolidation and hereditary creep in soils, {\it Comput. Methods Appl. Mech. Eng.} {\bf 190} (2001) 1985 - 2016.
	
\end{thebibliography}
\end{document}